\documentclass[aps, prd, showkeys, nofootinbib, floatfix]{revtex4-2}
\usepackage{amssymb}
\usepackage{amsmath}
\usepackage{graphicx}
\usepackage{dcolumn}
\usepackage{bm}
\usepackage{hyperref}
\begin{document}

\title{Third-order relativistic dissipative fluid dynamics from the method of moments}

\author{Caio V.~P.~de Brito}
\email{caio\_brito@id.uff.br}
\author{Gabriel S.~Denicol}
\email{gsdenicol@id.uff.br}
\affiliation{Instituto de F\'{\i}sica, Universidade Federal Fluminense, Niter\'oi, Rio de Janeiro, Brazil}

\begin{abstract}
We derive a linearly causal and stable third-order relativistic fluid-dynamical theory from the Boltzmann equation using the method of moments. For this purpose, we demonstrate that such theory must include novel degrees of freedom, corresponding to irreducible tensors of rank $3$ and $4$.
The equations of motion derived in this work are compared with numerical solutions of the Boltzmann equation, considering an ultrarelativistic, classical gas in the highly symmetric Bjorken flow scenario. These solutions are shown to be in good agreement for a wide range of values of shear viscosity and initial temperatures.
\end{abstract}

\maketitle

\section{Introduction}

Relativistic fluid dynamics has been successfully used to describe the quark-gluon plasma (QGP) created in relativistic heavy-ion collisions \cite{gale,heinz,monnai} at the Large Hadron Collider (LHC) and the Relativistic Heavy-ion Collider (RHIC). 
In order to accurately describe the evolution of the QGP, dissipative effects cannot be neglected as they are essential in describing rapidly expanding systems and, indeed, they play a fundamental role in describing the observed flow coefficients \cite{Romatschke_2007}. Currently, second-order theories of fluid dynamics are the most widely employed in the description of relativistic viscous fluids, since such theories can be constructed to be linearly casual and stable around global equilibrium \cite{hw1983, olson, denicolstability, pushi, bdcoupling, Sammet2023} \footnote{Recently, causal and stable generalizations of first-order fluid dynamics were extensively studied by Bemfica, Disconzi, Noronha and Kovtun \cite{bdn1, bdn2, kovtun1, kovtun2, bdn3}}. We remark that the most traditional second-order theory is due to Israel and Stewart \cite{is1}, developed in the 1970s for applications in cosmology, but several additional formulations have been developed ever since \cite{liu1986, carter1991, grmela1997, BRSSS, peralta2009, denicol2010, dnmr, dmnr, molnar2014}.

Naturally, one may also consider the derivation of third-order relativistic fluid-dynamical theories. As a matter of fact, several authors have already investigated this topic, using different frameworks e.g., a gradient expansion \cite{grozdanov}, a phenomenological description using the second law of thermodynamics \cite{el, muronga3} and kinetic theory, using a method inspired in the Chapman-Enskog expansion \cite{amaresh}. In particular, the latter formulation was shown to be in good agreement with solutions of the relativistic Boltzmann equation \cite{amaresh, amaresh0}. These studies were performed assuming the highly symmetric Bjorken flow scenario \cite{bjorken}, where solutions of the Boltzmann equation can actually be obtained without resorting to complex numerical schemes \cite{Florkowski_2013, El_2010}.

Recently, the third-order formalism developed in Ref.~\cite{amaresh} was shown to be linearly acausal and unstable \cite{bd3order}, presenting the same pathology originally observed in Navier-Stokes theory \cite{hw1985}. An \textit{ad hoc} modification to this theory was proposed in Ref.~\cite{bd3order}, in order to address the aforementioned problem. The goal of this paper is to obtain a more fundamental version of this framework from kinetic theory using the traditional method of moments \cite{dnmr}. We shall demonstrate that, in order to obtain equations that include all terms that are asymptotically of third order in gradients, it is necessary to include novel degrees of freedom that correspond to irreducible tensors of rank $3$ and $4$, while the fluid-dynamical theories developed so far only considered irreducible tensors of rank 0, 1, and 2. Finally, we show that solutions of our third-order fluid-dynamical formulation are in good agreement with solutions of the relativistic Boltzmann equation assuming a Bjorken flow scenario. 

This work is organized as follows: in Sec.~\ref{sec_fundamentals}, we discuss how the basic conservation laws emerge in a kinetic description and express all fluid-dynamical variables as moments of the single-particle distribution function. In Sec.~\ref{sec_mm}, we explain the method of moments and how the single-particle distribution function is expanded using a basis of irreducible tensors constructed from the 4-momenta. We also outline the equations of motion for the irreducible moments of the single-particle distribution function, which are used to derive fluid dynamics. In this section, we present, for the first time, the equations of motion for the irreducible moments of rank 3 and 4. In Sec.~\ref{sec_30mom}, we explain how fluid dynamics is derived using the method of moments and detail how the truncation procedure must be modified in order to obtain a transient third-order theory. Finally, in Sec.~\ref{sec_bjorken}, we compare the solutions of our theory with solutions of the relativistic Boltzmann equation in a Bjorken flow scenario. In Sec.~\ref{sec_conc}, we summarize our work and present our conclusions. Throughout this work, we adopt natural units, $c=k_B=\hbar=1$ and the mostly-minus convention for the Minkowski metric tensor, $g_{\mu\nu}=\mathrm{diag}\,(1,-1,-1,-1)$.

\section{Relativistic fluid dynamics and the Boltzmann equation}
\label{sec_fundamentals}

The Boltzmann equation describes the time evolution of the single-particle momentum distribution of a dilute gas. It is an integro-differential equation of the form, \cite{cercignani}
\begin{equation}
k^\mu \partial_\mu f_{\mathbf{k}}=\frac{1}{2}\int dK'dPdP' \left(f_{\mathbf{p}}f_{\mathbf{p}'}\tilde{f}_{\mathbf{k}}\tilde{f}_{\mathbf{k}'}-f_{\mathbf{k}}f_{\mathbf{k}'}\tilde{f}_{\mathbf{p}}\tilde{f}_{\mathbf{p'}}\right)\mathcal{W}_{\mathbf{k}\mathbf{k}'\leftrightarrow\mathbf{p}\mathbf{p}'}\equiv C[f], \label{boltz1}
\end{equation}
where $f_{\mathbf{k}} \equiv f(\mathbf{x},\mathbf{k})$ is the single-particle distribution function, $k^\mu=(\sqrt{\mathbf{k}^2+m^2},\mathbf{k})$ is the 4-momentum, with $m$ being the mass of the particles, $\mathcal{W}_{\mathbf{k}\mathbf{k}'\leftrightarrow\mathbf{p}\mathbf{p}'}$ is the Lorentz-invariant transition rate, and $dK=d\mathbf{k}/[(2\pi)^2k^0]$ is the Lorentz-invariant volume element in momentum space. We further defined $\tilde{f}_{\mathbf{k}}=1-af_{\mathbf{k}}$, with $a=-1$ ($a=1$) for bosons (fermions), while $a=0$ for classical particles.

The main equations of relativistic fluid dynamics are the continuity equations that describe the conservation of the number of particles (when considering binary collisions), energy and momentum, given respectively by
\begin{subequations}
\label{eq:cons_laws_short}
\begin{align}
\partial_\mu N^{\mu}&=0, \label{CNC}\\
\partial_\mu T^{\mu\nu}&=0, \label{CEM}
\end{align}
\end{subequations}
where $N^{\mu}$ is the particle 4-current and $T^{\mu\nu}$ is the energy-momentum tensor. These conserved currents can be expressed in terms of their parallel and orthogonal components with respect to the fluid 4-velocity, $u^\mu = \gamma(1, \mathbf{V})$, a normalized timelike 4-vector, $u_\mu u^\mu = 1$, with $\gamma = 1/\sqrt{1-V^2}$ being the Lorentz factor. In this case, they read
\begin{subequations}
\begin{align}
N^\mu&=nu^\mu+n^\mu, \label{nmu_u} \\
T^{\mu\nu}&=\varepsilon u^\mu u^\nu-\Delta^{\mu\nu}(P_0+\Pi)+W^\mu u^\nu+W^\nu u^\mu+\pi^{\mu\nu}, \label{tmunu_u}
\end{align}
\end{subequations}
where $n$ is the particle density in the local rest frame of the fluid, $n^\mu$ is the particle diffusion 4-current, $\varepsilon$ is the energy density in the local rest frame of the fluid, $\Delta^{\mu\nu} = g^{\mu\nu} - u^\mu u^\nu$ is the orthogonal projection operator onto the 3-space orthogonal to $u^\mu$, $P_0$ is the thermodynamic pressure, $\Pi$ is the bulk viscous pressure, $W^\mu$ is the energy diffusion 4-current, and $\pi^{\mu\nu}$ is the shear-stress tensor. 

The conservation laws, Eqs.~\eqref{eq:cons_laws_short}, can also be decomposed in terms of parallel and orthogonal components with respect to $u^\mu$, leading to
\begin{subequations}
\label{eqs:cons_laws}
\begin{align}
\partial_\mu N^\mu &= \dot{n} + n\theta + \nabla_\mu n^\mu = 0, \label{charge_cons}  \\
u_{\nu }\partial _{\mu }T^{\mu \nu } &= \dot{\varepsilon}+\,\left(
\varepsilon +P\right) \theta -\pi ^{\alpha \beta }\sigma _{\alpha \beta }=0,
\label{energy_cons} \\
\Delta _{\nu }^{\lambda }\partial _{\mu }T^{\mu \nu } &= \left( \varepsilon
+P\right) \dot{u}^{\lambda }-\nabla ^{\lambda }P-\pi ^{\lambda \beta }\dot{u}%
_{\beta }+\Delta _{\nu }^{\lambda }\nabla _{\mu }\pi ^{\mu \nu }=0,
\label{momentum_cons}
\end{align}
\end{subequations}
where $u^\mu \partial_\mu = d/d\tau$ is the comoving time derivative, $\theta = \partial_\mu u^\mu$ is the expansion rate, $\sigma^{\mu\nu} = \Delta^{\mu\nu\alpha\beta} \nabla_\alpha u_\beta$ is the shear tensor, with $\Delta^{\mu\nu\alpha\beta} = (\Delta^{\mu\alpha}\Delta^{\nu\beta}+\Delta^{\mu\beta}\Delta^{\nu\alpha})/2 - \Delta^{\mu\nu}\Delta^{\alpha\beta}/3$ being the double, symmetric and traceless projection operator orthogonal to the 4-velocity, and $\nabla_\mu = \Delta^\nu_\mu \partial_\nu$ is the projected  4-derivative.

In the context of relativistic kinetic theory, the particle 4-current and the energy-momentum tensor are identified as the first and second moments of the single-particle distribution function, respectively,  \cite{degroot, Denicol:2021}
\begin{subequations}
\begin{align}
N^\mu&=\int dK k^\mu f_\mathbf{k}\equiv\left\langle k^\mu\right\rangle, \label{nmu_moment}\\
T^{\mu\nu}&=\int dK k^\mu k^\nu f_\mathbf{k}\equiv\left\langle k^\mu k^\nu\right\rangle. \label{tmunu_moment}
\end{align}
\end{subequations}
Once again, following a decomposition in terms of $u^\mu$, the 4-momentum, $k^\mu$, can be expressed as $k^\mu = E_{\mathbf{k}} u^\mu + k^{\langle\mu\rangle}$, where $E_{\mathbf{k}} \equiv u_\mu k^\mu$ is the energy in the local rest frame of the fluid, and $k^{\langle\mu\rangle} \equiv \Delta^{\mu\nu}k_\nu$ is the orthogonal projection of the 4-momentum. Then, Eqs.~\eqref{nmu_moment} and \eqref{tmunu_moment} can be cast in the following form
\begin{subequations}
\begin{align}
N^\mu&=\left\langle E_{\mathbf{k}}\right\rangle u^\mu+\left\langle k^{\langle\mu\rangle}\right\rangle, \label{nmu_int} \\
T^{\mu\nu}&= \left\langle E_{\mathbf{k}}^2 \right\rangle u^\mu u^\nu+\frac{1}{3}\Delta^{\mu\nu}\langle b_{\mathbf{k}} \rangle+\left\langle E_{\mathbf{k}} k^{\langle\mu\rangle}\right\rangle u^\nu+\left\langle E_{\mathbf{k}} k^{\langle\nu\rangle}\right\rangle u^\mu+\left\langle k^{\langle\mu}k^{\nu\rangle}\right\rangle, \label{tmunu_int}
\end{align}
\end{subequations}
where we have defined $b_{\mathbf{k}} \equiv \Delta_{\mu\nu}k^\mu k^\nu$ and employed the notation $A^{\langle\mu_1\cdots\mu_\ell\rangle} \equiv \Delta^{\mu_1\cdots\mu_\ell}_{\nu_1\cdots\nu_\ell} A^{\nu_1\cdots\nu_\ell}$ to denote the irreducible projection of an arbitrary tensor $A^{\nu_1\cdots\nu_\ell}$, with $\Delta^{\mu_1\cdots\mu_\ell}_{\nu_1\cdots\nu_\ell}$ being the $2\ell$-index traceless (for $\ell >1$) and symmetric projection operator orthogonal to the 4-velocity in all $\mu$- and $\nu$-type indices, cf.~Refs.~\cite{degroot, Denicol:2021}. Comparing Eqs.~\eqref{nmu_int} and \eqref{tmunu_int} to Eqs.~\eqref{nmu_u} and \eqref{tmunu_u}, respectively, it is then possible to identify the fluid-dynamical fields as moments of $f_{\mathbf{k}}$,
\begin{equation}
n=\left\langle E_{\mathbf{k}}\right\rangle, \hspace{.2cm} n^\mu=\left\langle k^{\langle\mu\rangle}\right\rangle, \hspace{.2cm} \varepsilon=\langle E_{\mathbf{k}}^2\rangle, \hspace{.2cm} P_0+\Pi=-\frac{1}{3}\langle b_{\mathbf{k}} \rangle, \hspace{.2cm} W^\mu=\left\langle E_{\mathbf{k}} k^{\langle\mu\rangle}\right\rangle, \hspace{.2cm} \pi^{\mu\nu}=\left\langle k^{\langle\mu}k^{\nu\rangle}\right\rangle. \label{quantities1}
\end{equation}

It is then convenient to express the single-particle distribution function as
\begin{equation}
f_\mathbf{k}=f_{0\mathbf{k}}+\delta f_\mathbf{k} \label{expand_f},
\end{equation}
where $f_{0\mathbf{k}}$ is the local equilibrium distribution function,
\begin{equation}
f_{0\mathbf{k}}=\frac{1}{\exp({\beta_0 E_{\mathbf{k}}-\alpha_0})+a},
\end{equation}
with $\beta_0 = 1/T$ being the inverse temperature, $\alpha_0 = \mu/T$ being the chemical potential over temperature, commonly referred to as the thermal potential, and $\delta f_\mathbf{k}$ denoting a deviation from equilibrium. At this point, it is convenient to employ the following notation, introduced in Ref.~\cite{dnmr},
\begin{equation}
\langle \cdots\rangle\equiv\langle \cdots\rangle_0+\langle \cdots\rangle_\delta, \hspace{.2cm}\text{where}\left\{ 
\begin{array}{cc}
\langle \cdots\rangle_0\equiv\int dK (\cdots)f_{0\mathbf{k}}, \\ 
\langle \cdots\rangle_\delta\equiv\int dK (\cdots)\delta f_{\mathbf{k}}.
\end{array}
\right. 
\end{equation}
Wherefore, from Eq.~\eqref{quantities1}, we can write the equilibrium and nonequilibrium fluid-dynamical fields as
\begin{equation}
n_0\equiv\left\langle E_{\mathbf{k}}\right\rangle_0,\text{ } n^\mu=\left\langle k^{\langle\mu\rangle}\right\rangle_\delta, \hspace{.1cm} \varepsilon_0\equiv\langle E_{\mathbf{k}}^2\rangle_0, \hspace{.1cm} P_0=-\frac{1}{3}\langle b_{\mathbf{k}} \rangle_0, \hspace{.1cm} \Pi=-\frac{1}{3}\langle b_{\mathbf{k}} \rangle_\delta, \hspace{.1cm} W^\mu=\left\langle E_{\mathbf{k}} k^{\langle\mu\rangle}\right\rangle_\delta, \hspace{.1cm} \pi^{\mu\nu}=\left\langle k^{\langle\mu}k^{\nu\rangle}\right\rangle_\delta. \label{quantities2}
\end{equation}
Here, matching conditions are employed to determine the temperature, $T$, and chemical potential, $\mu$, in such a way that the particle density and the energy density are fixed to their respective equilibrium values, thus leading to the conditions $\langle E_{\mathbf{k}}\rangle_\delta=\langle E_{\mathbf{k}}^2\rangle_\delta=0$ \cite{Denicol:2021}. In the following, we shall use Landau matching conditions \cite{landau} which further impose that the energy diffusion 4-current vanishes, $W^\mu = 0$.

\section{Method of moments}
\label{sec_mm}

In this section we provide a brief introduction of the method of moments \cite{degroot, Denicol:2021, dnmr}, originally developed by Grad for nonrelativistic systems \cite{Grad}. It is one of the most widespread frameworks to derive relativistic fluid dynamics from the Boltzmann equation. Unlike the Chapman-Enskog method \cite{chapman1970mathematical}, which yields theories that are unsuitable to describe relativistic fluids \cite{hw1985, pushi, bdcoupling, denicolstability, Denicol:2016bjh, nonlinear1, nonlinear2, Gavassino_2022}, the method of moments leads to formulations that may be causal and stable, at least in the linear regime \cite{hw1983, olson, denicolstability, pushi, bdcoupling, Sammet2023}.

\subsection{Outline}

The first step is to rewrite Eq.~\eqref{expand_f} in the following form, 
\begin{equation}
f_\mathbf{k}=f_{0\mathbf{k}} \left( 1 + \tilde{f}_{0\mathbf{k}} \phi_\mathbf{k} \right), \label{expand_f2}
\end{equation}
where we have expressed the nonequilibrium correction to the single-particle distribution function as $\phi_\mathbf{k} \equiv \delta f_\mathbf{k}/(f_{0\mathbf{k}}\tilde{f}_{0\mathbf{k}})$.
Then, $\phi_{\mathbf{k}}$ itself is expanded in terms of a complete orthogonal basis of irreducible tensors,
\begin{equation}
\phi_{\mathbf{k}}=\sum_{\ell=0}^\infty \lambda_{\mathbf{k}}^{\langle\mu_1\cdots\mu_\ell\rangle}k_{\langle\mu_1}\cdots k_{\mu_\ell\rangle}, \label{phi_series}
\end{equation}
where the tensors $k^{\langle\mu_1}\cdots k^{\mu_\ell\rangle}$ satisfy the following orthogonality condition \cite{degroot,Denicol:2021},
\begin{equation}
\int dK F(E_{\mathbf{k}}) k^{\langle\mu_1}\cdots k^{\mu_\ell\rangle}k_{\langle\nu_1}\cdots k_{\nu_m\rangle}=\frac{\ell!\delta_{\ell m}}{(2\ell+1)!!}\Delta^{\mu_1\cdots\mu_\ell}_{\nu_1\cdots\nu_m}\int dK F(E_{\mathbf{k}}) b_{\mathbf{k}}^m, \label{orthogonality_k}
\end{equation}
with $F(E_{\mathbf{k}})$ being an arbitrary function of $E_{\mathbf{k}}$ \cite{dnmr}. The expansion coefficients $\lambda_{\mathbf{k}}^{\langle \mu_1 \cdots \mu_\ell \rangle}$ can be further expanded using a basis of orthogonal functions $P^{(\ell)}_{\mathbf{k}n}$, 
\begin{equation}
\lambda_{\mathbf{k}}^{\langle\mu_1\cdots\mu_\ell\rangle}=\sum_{n=0}^{N_\ell}\Phi_n^{\langle\mu_1\cdots\mu_\ell\rangle}P^{(\ell)}_{\mathbf{k}n}, \label{def_lambda1}
\end{equation}
with $N_\ell$ being the number of terms considered in the expansion. A detailed discussion on how to truncate this expansion is developed in Sec.~\ref{sec_30mom}. The functions $P^{(\ell)}_{\mathbf{k}n}$ are conveniently chosen to be a power series of $E_{\mathbf{k}}$, 
\begin{equation}
P^{(\ell)}_{\mathbf{k}n}=\sum_{r=0}^n a^{(\ell)}_{nr}E_{\mathbf{k}}^r, \label{def_P}
\end{equation}
and are constructed to satisfy the following orthogonality condition
\begin{equation}
\int dK \omega^{(\ell)}P^{(\ell)}_{\mathbf{k}n}P^{(\ell)}_{\mathbf{k}m}=\delta_{mn},  \label{ortho1}
\end{equation}
with the weight $\omega^{(\ell)}$ being
\begin{equation}
\omega^{(\ell)}=\frac{\mathcal{N}^{(\ell)}}{(2\ell+1)!!}\left(\Delta_{\alpha\beta}k^\alpha k^\beta\right)^\ell f_{0\mathbf{k}}\tilde{f}_{0\mathbf{k}}.
\end{equation}
For the sake of convenience, we take $P^{(\ell)}_{\mathbf{k}0}=1$, for any value of $\ell$. The remaining elements of the basis, as well as the normalization constants $\mathcal{N}^{(\ell)}$, are then obtained using the Gram-Schmidt orthogonalization procedure, see Appendix E of Ref.~\cite{dnmr} and Ref.~\cite{Denicol:2021} for details.

The irreducible moments of the nonequilibrium distribution function are defined as
\begin{equation}
\rho^{\mu_1\cdots\mu_\ell}_r\equiv\left\langle E^r_{\mathbf{k}}k^{\langle\mu_1}\cdots k^{\mu_\ell\rangle}\right\rangle_\delta, \label{def_rhos}
\end{equation}
and can be shown to be related to $\Phi^{\langle\mu_1\cdots\mu_\ell\rangle}_n$ through
\begin{equation}
\Phi^{\langle\mu_1\cdots\mu_\ell\rangle}_n = \frac{\mathcal{N}^{(\ell)}}{\ell!}\sum_{r=0}^{n} \rho^{\mu_1\cdots\mu_\ell}_r a_{nr}^{(\ell)}. \label{def_Phi}
\end{equation}
Finally, the expansion of the single-particle distribution in a basis of irreducible momenta reads
\begin{equation}
f_{\mathbf{k}} =  f_{0 \mathbf{k}} + f_{0 \mathbf{k}}\tilde{f}_{0\mathbf{k}} \sum_{\ell = 0}^\infty \sum_{n = 0}^{N_\ell} \sum_{r=0}^{n} \frac{\mathcal{N}^{(\ell)}}{\ell!} a^{(\ell)}_{n r} \, P_{\mathbf{k} n}^{(\ell)} \, \rho^{\mu_1 \cdots \mu_\ell}_r \, k_{\langle \mu_1} \cdots k_{\mu_\ell \rangle} . \label{eq:usual_moments}
\end{equation}

At this point, we note that the fluid-dynamical variables in Eqs.~\eqref{quantities2} can be expressed in terms of the irreducible moments as
\begin{equation}
\Pi=-\frac{1}{3}m^2\rho_0, \hspace{.1cm} n^\mu=\rho^\mu_0, \hspace{.1cm} W^\mu=\rho^\mu_1, \hspace{.1cm} \pi^{\mu\nu}=\rho^{\mu\nu}_0. \label{quantities_rho}
\end{equation}
As it was previously discussed, the traditional Landau matching conditions imply that $\rho_1=0$, $\rho_2=0$, and $W^\mu=\rho^\mu_1=0$. We note that a derivation of fluid dynamics from the method of moments considering generic matching conditions was studied in Ref.~\cite{gsr}.

\subsection{Equations of motion}

In order to obtain the evolution of the dissipative currents, it is necessary to calculate the equations of motion satisfied by the irreducible moments. The equations of motion for the moments of rank 0, 1, and 2 have already been derived in Ref.~\cite{dnmr} and are reproduced below. First, the equation of motion for the scalar moment is
\begin{eqnarray}
\dot{\rho}_r&=&C_{r-1}+\alpha_r^{(0)}\theta+r\rho_{r-1}^\mu\dot{u}_\mu-\nabla_\mu \rho^\mu_{r-1}+\frac{G_{3r}}{D_{20}}\partial_\mu n^\mu+\left[(r-1)\rho_{r-2}^{\mu\nu}+\frac{G_{3r}}{D_{20}}\pi^{\mu\nu}\right]\sigma_{\mu\nu} \notag \\
&+&\frac{1}{3}\left[(r-1)m^2\rho_{r-2}-(r+2)\rho_r-3\frac{G_{2r}}{D_{20}}\Pi\right]\theta. \label{eom0}
\end{eqnarray}
The moments of rank 1 satisfy the following equation of motion
\begin{eqnarray}
\dot{\rho}^{\langle\mu\rangle}_r&=&C^{\langle\mu\rangle}_{r-1}+\alpha^{(1)}_r\nabla^\mu\alpha_0+r\dot{u}_\nu\rho^{\mu\nu}_{r-1}-\Delta^\mu_\alpha\nabla_\beta\rho^{\alpha\beta}_{r-1}+\rho^{\langle\nu}_r\omega^{\mu\rangle}{}_\nu+\frac{1}{3}\left[(r-1)m^2\rho^\mu_{r-2}-(r+3)\rho_r^\mu\right]\theta+(r-1)\sigma_{\alpha\beta}\rho^{\mu\alpha\beta}_{r-2}+\notag\\
&+&\frac{1}{3}\left[rm^2\rho_{r-1}-(r+3)\rho_{r+1}\right]\dot{u}^\mu-\frac{1}{3}\nabla^\mu\left(m^2\rho_{r-1}-\rho_{r+1}\right)+\frac{1}{5}\left[(2r-2)m^2\rho^\nu_{r-2}-(2r+3)\rho^\nu_r\right]\sigma^\nu_\mu+\notag\\
&+&\frac{\beta_0 J_{r+2,1}}{\varepsilon_0+P_0}\left(\Pi\dot{u}^\mu-\nabla^\mu \Pi+\Delta^\mu_\nu\partial_\lambda \pi^{\nu\lambda}\right), \label{eom1}
\end{eqnarray}
while the irreducible moments of rank 2 satisfy
\begin{eqnarray}
\dot{\rho}_r^{\langle\mu\nu\rangle}&=&C^{\langle\mu\nu\rangle}_{r-1}+2\alpha^{(2)}_r\sigma^{\mu\nu}+\frac{2}{15}\left[(r-1)m^4\rho_{r-2}-(2r+3)m^2\rho_r+(r+4)\rho_{r+2}\right]\sigma^{\mu\nu}+r\rho^{\mu\nu\lambda}_{r-1}\dot{u}_\lambda+\notag \\
&+&\frac{2}{5}\left[rm^2\rho^{\langle\mu}_{r-1}-(r+5)\rho^{\langle\mu}_{r+1}\right]\dot{u}^{\nu\rangle}-\frac{2}{5}\nabla^{\langle\mu}\left(m^2\rho^{\nu\rangle}_{r-1}-\rho^{\nu\rangle}_{r+1}\right)+2\rho^{\lambda\langle\mu}_r\omega^{\nu\rangle}{}_\lambda+(r-1)\rho^{\mu\nu\alpha\beta}_{r-2}\sigma_{\alpha\beta}+\notag \\
&+&\frac{2}{7}\left[(2r-2)m^2\rho^{\lambda\langle\mu}_{r-2}-(2r+5)\rho^{\lambda\langle\mu}_{r}\right]\sigma^{\nu\rangle}_\lambda+\frac{1}{3}\left[(r-1)m^2\rho^{\mu\nu}_{r-2}-(r+4)\rho^{\mu\nu}_r\right]\theta-\Delta^{\mu\nu}_{\alpha\beta}\nabla_\lambda\rho^{\alpha\beta\lambda}_{r-1}. \label{eom2}
\end{eqnarray}

Furthermore, in this work, we also calculate the equations of motion for the irreducible moments of rank 3 and 4, since these currents are essential to the derivation of a third-order formalism \cite{bd3order}, as it will be discussed later. For the irreducible moments of rank 3, we obtain
\begin{eqnarray}
\dot{\rho}^{\langle\mu\nu\lambda\rangle}_r&=&C^{\langle\mu\nu\lambda\rangle}_{r-1}+\frac{1}{3}\left[(r-1)m^2\rho^{\mu\nu\lambda}_{r-2}-(r+5)\rho^{\mu\nu\lambda}_{r}\right]\theta+\frac{6}{35}\sigma^{\langle\mu\nu}\left[(r-1)m^4\rho^{\lambda\rangle}_{r-2}-(2r+5)m^2\rho^{\lambda\rangle}_r+(r+6)\rho^{\lambda\rangle}_{r+2}\right]+\notag\\
&+&3\rho^{\alpha\langle\mu\nu}_r\omega^{\lambda\rangle}{}_\alpha+\frac{1}{3}\sigma^{\langle\mu}_\alpha\left[m^2(2r-2)\rho^{\nu\lambda\rangle\alpha}_{r-2}-(2r+7)\rho^{\nu\lambda\rangle\alpha}_r\right]+r\dot{u}_\alpha\rho^{\mu\nu\lambda\alpha}-\frac{3}{7}\nabla^{\langle\mu}\left(m^2\rho^{\nu\lambda\rangle}_{r-1}-\rho^{\nu\lambda\rangle}_{r+1}\right)+ \label{eom3}\\
&+&\frac{3}{7}\left[rm^2\rho^{\langle\mu\nu}_{r-1}-(r+7)\rho^{\langle\mu\nu}_{r+1}\right]\dot{u}^{\lambda\rangle}-\Delta^{\mu\nu\lambda}_{\alpha\beta\sigma}\nabla_\gamma\rho^{\alpha\beta\sigma\gamma}_{r-1}+(r-1)\sigma_{\alpha\beta}\rho^{\mu\nu\lambda\alpha\beta}_{r-2}. \notag
\end{eqnarray}
Finally, the irreducible moments of rank 4 satisfy the following equations of motion
\begin{eqnarray}
\dot{\rho}^{\langle\mu\nu\alpha\beta\rangle}_r&=&C^{\langle\mu\nu\alpha\beta\rangle}_{r-1}+r\dot{u}_\lambda\rho^{\mu\nu\alpha\beta\lambda}_r+4\dot{u}^{\langle\mu}\left[\frac{rm^2}{9}\rho^{\nu\alpha\beta\rangle}_{r-1}-\left(1+\frac{r}{9}\right)\rho^{\nu\alpha\beta\rangle}_{r+1}\right]-\frac{4}{9}\nabla^{\langle\mu}\left(m^2\rho^{\nu\alpha\beta\rangle}_{r-1}-\rho^{\nu\alpha\beta\rangle}_{r+1}\right)-\Delta^{\mu\nu\alpha\beta}_{\sigma\gamma\psi\rho}\nabla_\lambda\rho^{\sigma\gamma\psi\rho\lambda}_{r-1}+\notag\\
&+&\frac{4}{21}\sigma^{\langle\mu\nu}\left[(r-1)m^4\rho^{\alpha\beta\rangle}_{r-2}-(2r+7)m^2\rho^{\alpha\beta\rangle}_r+(r+8)\rho^{\alpha\beta\rangle}_{r+2}\right]+4\rho^{\lambda\langle\mu\nu\alpha}_r\omega^{\beta\rangle}{}_\lambda+(r-1)\sigma_{\lambda\sigma}\rho^{\mu\nu\alpha\beta\lambda\sigma}_{r-2}+\notag\\
&+&\frac{4}{11}\sigma^{\langle\mu}_\lambda\left[(2r-2)m^2\rho^{\nu\alpha\beta\rangle\lambda}_{r-2}-(2r+9)\rho^{\nu\alpha\beta\rangle\lambda}_r\right]+\frac{1}{3}\left[(r-1)m^2\rho^{\mu\nu\alpha\beta}_{r-2}-\left(r+6\right)\rho^{\mu\nu\alpha\beta}_r\right]\theta. \label{eom4}
\end{eqnarray}
In deriving these equations, we have used the following identities
\begin{subequations}
\begin{align}
k^{\langle\mu\rangle}k^{\langle\nu\rangle}&=k^{\langle\mu}k^{\nu\rangle}+\frac{1}{3}\Delta^{\mu\nu}b_{\mathbf{k}},\\
k^{\langle\mu\rangle}k^{\langle\nu\rangle}k^{\langle\alpha\rangle}&=k^{\langle\mu}k^{\nu}k^{\alpha\rangle}+\frac{1}{5}b_{\mathbf{k}}\Delta^{(\mu\nu}k^{\langle\alpha\rangle)},\\
k^{\langle\mu\rangle}k^{\langle\nu\rangle}k^{\langle\alpha\rangle}k^{\langle\beta\rangle}&=k^{\langle\mu}k^{\nu}k^{\alpha}k^{\beta\rangle}+\frac{6}{7}b_{\mathbf{k}}\Delta^{(\mu\nu}k^{\langle\alpha\rangle}k^{\langle\beta\rangle)}-\frac{3}{35}b_{\mathbf{k}}\Delta^{\mu(\nu}\Delta^{\alpha\beta)},\\
k^{\langle\mu\rangle}k^{\langle\nu\rangle}k^{\langle\alpha\rangle}k^{\langle\beta\rangle}k^{\langle\rho\rangle}&=k^{\langle\mu}k^{\nu}k^{\alpha}k^{\beta}k^{\rho\rangle}+\frac{10}{9}b_{\mathbf{k}}\Delta^{(\mu\nu}k^{\langle\alpha\rangle}k^{\langle\beta\rangle}k^{\langle\rho\rangle)}-\frac{15}{63}b_{\mathbf{k}}^2 \Delta^{(\mu\nu}\Delta^{\alpha\beta}k^{\langle\rho\rangle)},\\
k^{\langle\mu\rangle}k^{\langle\nu\rangle}k^{\langle\alpha\rangle}k^{\langle\beta\rangle}k^{\langle\rho\rangle}k^{\langle\sigma\rangle}&=k^{\langle\mu}k^{\nu}k^{\alpha}k^{\beta}k^{\rho}k^{\sigma\rangle}+\frac{15}{11}b_{\mathbf{k}}\Delta^{(\mu\nu}k^{\langle\alpha\rangle}k^{\langle\beta\rangle}k^{\langle\rho\rangle}k^{\langle\sigma\rangle)}-\frac{45}{99}b_\mathbf{k}^2\Delta^{(\mu\nu}\Delta^{\alpha\beta}k^{\langle\rho\rangle}k^{\langle\sigma\rangle)}\\
&+\frac{15}{693}b^3_{\mathbf{k}}\Delta^{(\mu\nu}\Delta^{\alpha\beta}\Delta^{\rho\sigma)}, \notag
\end{align}
\end{subequations}
where the parentheses denote all possible permutations between the indices. We further used the equations of motion for the thermal potential, inverse temperature and 4-velocity, that stem from the conservation laws, Eqs.~\eqref{eqs:cons_laws},
\begin{subequations}
\begin{align}
\dot{\alpha}_0&=\frac{1}{D_{20}}\left\{-J_{30}\left(n_0\theta+\partial_\mu n^\mu\right)+J_{20}\left[\left(\varepsilon_0+P_0+\Pi\right)\theta-\pi^{\mu\nu}\sigma_{\mu\nu}\right]\right\},\\
\dot{\beta}_0&=\frac{1}{D_{20}}\left\{-J_{20}\left(n_0\theta+\partial_\mu n^\mu\right)+J_{10}\left[\left(\varepsilon_0+P_0+\Pi\right)\theta-\pi^{\mu\nu}\sigma_{\mu\nu}\right]\right\},\\
\dot{u}^\mu&=\frac{1}{\varepsilon_0+P_0}\left(\nabla^\mu P-\Pi\dot{u}^\mu+\nabla^\mu\Pi-\Delta^\mu_\nu\partial_\lambda\pi^{\nu\lambda}\right), \label{eq:udot}
\end{align}
\end{subequations}
and introduced the following thermodynamic variables
\begin{subequations}
\begin{align}
\alpha_r^{(0)}&\equiv(1-r)I_{r1}-I_{r0}-\frac{n_0}{D_{20}}\left(h_0G_{2r}-G_{3r}\right),\\
\alpha^{(1)}_r&\equiv J_{r+1,1}-h_0^{-1}J_{r+2,1},\\
\alpha_r^{(2)}&\equiv I_{r+2,1}+(r-1)I_{r+2,2}.
\end{align}
\end{subequations}
We have also employed the notation
\begin{subequations}
\begin{align}
I_{ij}&=\frac{(-1)^j}{(2j+1)!!}\int dK E^{i-2j}_{\mathbf{k}} b_{\mathbf{k}}^j f_{0\mathbf{k}},\\
J_{ij}&=\frac{(-1)^j}{(2j+1)!!}\int dK E^{i-2j}_{\mathbf{k}} b_{\mathbf{k}}^j f_{0\mathbf{k}}\tilde{f}_{0\mathbf{k}},\\
G_{ij}&=J_{i0}J_{j0}-J_{i-1,0}J_{j+1,0},\\
D_{ij}&=J_{i+1,j}J_{i-1,j}-(J_{ij})^2,
\end{align}
\end{subequations}
and defined the generalized collision term
\begin{equation}
C^{\langle\mu_1\cdots\mu_\ell\rangle}_r\equiv\int dK E^r_{\mathbf{k}}k^{\langle\mu_1}\cdots k^{\mu_\ell\rangle}C[f],
\end{equation}
following the notation of Ref.~\cite{dnmr}.

The equations of motion for the irreducible moments up to rank 2, Eqs.~\eqref{eom0}-\eqref{eom2}, all have the same structure: the dominant terms are the so-called Navier-Stokes terms, which are given by irreducible projections of first order derivatives of temperature, chemical potential, and 4-velocity, i.e., $\nabla^\mu \alpha$, $\theta$ and $\sigma^{\mu\nu}$. These type of terms are of first order in Knudsen number and appear as the dominant contribution in a gradient expansion of these irreducible tensors \cite{degroot, Denicol:2021}. The remaining terms in these equations of motion are at least of second order in Knudsen number in a gradient expansion. 

The equations of motion for the irreducible moments of rank 3 and 4, Eqs.~\eqref{eom3} and \eqref{eom4}, respectively, on the other hand, display qualitative differences from the scenario depicted above. First, such equations do not contain any term that is of first order in Knudsen number. This happens because it is not possible to construct irreducible tensors of rank higher than 2 solely from first order derivatives of temperature, chemical potential, and 4-velocity. Therefore, in these equations of motion, the dominant contribution in a gradient expansion is at least of second order in Knudsen number. Thus, in the same way that gradients of $T$, $\mu$ and $u^\mu$ act as source terms for the dissipative currents appearing in $N^\mu$ and $T^{\mu\nu}$, irreducible moments of rank 1 and 2 and derivatives thereof act as the dominant source terms for the irreducible moments of rank 3 and 4. 

We finally note that the irreducible moments of rank 3 and 4 only appear in the equations of motion for $\rho_r$, $\rho^\mu_r$ and $\rho_r^{\mu\nu}$ multiplied by a term of first order in gradients or being differentiated. Thus, such contributions would be at least of third order in a gradient expansion. This is the reason why these terms are usually neglected in the derivation of the so-called second-order theories \cite{dnmr}, since such formulations only include contributions to the conserved currents that are up to second order in a gradient expansion. In this work, our goal is to derive a third-order theory and, for this purpose, these contributions cannot be ignored. Thus, as argued in Ref.~\cite{bd3order}, we shall include degrees of freedom that can be matched to irreducible tensors of rank 3 and 4 and will incorporate such corrections. We remark that irreducible moments of rank 5 or higher are at least of third order in Knudsen number in a gradient expansion (appearing as corrections of fourth order or higher in the dynamics of the particle diffusion 4-current and the shear-stress tensor) and will not contribute to a third-order formulation. 

The next step is to use the equations of motion for the irreducible moments of the nonequilibrium distribution function, Eqs.~\eqref{eom0}--\eqref{eom4}, to obtain a closed set of equations of motion for the dissipative currents. This task is the main goal of this paper and will be carefully performed in the following section.

\section{Hydrodynamic equations from the method of moments}
\label{sec_30mom}

It is now necessary to truncate the expansion of the nonequilibrium distribution function in order to describe a fluid-dynamical system using a reduced -- and, in particular, \textit{finite} -- number of degrees of freedom. In this section, we detail the truncation procedure adopted in this work, an extension of the 14-moment approximation developed by Israel and Stewart \cite{is1}, and employ it to obtain a set of third-order fluid-dynamical equations.

\subsection{A new minimal truncation scheme}

In their original work, Israel and Stewart \cite{is1} truncated the expansion of $\phi_{\mathbf{k}}$ at second order in momenta,
\begin{equation}
\phi^{\mathrm{IS}}_{\mathbf{k}} = \lambda + \lambda^\mu k_\mu + \lambda^{\mu\nu} k_\mu k_\nu + \mathcal{O}(k^3). \label{eq:phi_DNMR}
\end{equation}
In this truncated expansion, there is a total of 14 degrees of freedom, which can be matched to the number of independent components of the particle 4-current, $N^\mu$, and the energy-momentum tensor, $T^{\mu\nu}$. This procedure is usually referred to as the 14-moment approximation. 

This approach can also be implemented using the \textit{complete} basis of irreducible tensors introduced in the previous section \cite{dnmr}. In this case, one expresses $\phi_{\mathbf{k}}$ up to second order in momenta as
\begin{equation}
\phi_{\mathbf{k}}^{\mathrm{IS}} = \lambda_{\mathbf{k}} + \lambda_{\mathbf{k}}^{\langle\mu\rangle} k_{\langle\mu\rangle} + \lambda^{\langle\mu\nu\rangle}_{\mathbf{k}} k_{\langle\mu} k_{\nu\rangle} + \mathcal{O}(k^3).
\end{equation}
The coefficients $\lambda_{\mathbf{k}}$, $\lambda_{\mathbf{k}}^{\langle\mu\rangle}$, and $\lambda^{\langle\mu\nu\rangle}_{\mathbf{k}}$ now carry a momentum dependence and are written in terms of orthogonal polynomials so that only terms that are of second-order or less in momentum remain, 
\begin{subequations}
\label{eq:lambda_DNMR}
\begin{align}
\lambda_{\mathbf{k}}&=\Phi_0+P^{(0)}_{\mathbf{k}1}\Phi_1+P^{(0)}_{\mathbf{k}2}\Phi_2,\\
\lambda_{\mathbf{k}}^{\langle\mu\rangle}&=\Phi^{\langle\mu\rangle}_0+P^{(1)}_{\mathbf{k}1}\Phi^{\langle\mu\rangle}_1,\\
\lambda_{\mathbf{k}}^{\langle\mu\nu\rangle}&=\Phi^{\langle\mu\nu\rangle}_0.
\end{align}
\end{subequations}
This approximation corresponds to truncating the expansion defined in \eqref{def_lambda1} using $N_0 = 2$, $N_1 = 1$, and $N_2 = 0$ and is equivalent to the 14-moment approximation proposed by Israel and Stewart. 

As already stated, the truncation above is not motivated by an expansion in terms of a small parameter, as occurs in the Chapman-Enskog expansion \cite{chapman1970mathematical}. It is a truncation in degrees of freedom and one simply stops when the number of degrees of freedom in the expansion becomes identical to the number of degrees of freedom expected in the fluid-dynamical theory (in the case of second-order fluid dynamics, 14 fields). For this reason, we included three terms in the expansion of the scalar coefficient ($\ell=0$), since one of them is mapped onto the bulk viscous pressure, while the other two are determined from the matching conditions that define the temperature and chemical potential. For the 4-vector coefficient ($\ell=1$), we included two terms, since one is mapped onto the particle diffusion 4-current and the other onto the energy diffusion 4-current -- one of these currents (here, the energy diffusion) is traditionally eliminated by matching conditions that define the fluid 4-velocity. Finally, for $\ell=2$, we included only one term in the expansion, that is mapped onto the shear-stress tensor. This truncation procedure is usually referred to as a \textit{minimal truncation scheme}.

Recently, it was argued in Ref.~\cite{bd3order} that, in order to obtain a linearly causal and stable third-order fluid-dynamical theory, the shear-stress tensor must be coupled to a novel dissipative current that satisfies its own transient dynamics. In the linear regime, this novel degree of freedom corresponds to an irreducible third-rank tensor, which effectively describes part of the third-order contributions that appear in the theory. As we argued in the previous section, in the nonlinear regime, fourth-rank tensors will also contribute to a third-order fluid-dynamical theory. Therefore, this will require including more degrees of freedom in the moment expansion and the truncation procedure described above for the nonequilibrium correction $\phi_\mathbf{k}$ must be reevaluated. 

We then propose a new minimal truncation for the expansion of $\phi_{\mathbf{k}}$ that takes the following form
\begin{equation}
\phi_{\mathbf{k}}^{\mathrm{3rd-order}} = \lambda_{\mathbf{k}} + \lambda_{\mathbf{k}}^{\langle\mu\rangle} k_{\langle\mu\rangle}+\lambda^{\langle\mu\nu\rangle}_{\mathbf{k}} k_{\langle\mu} k_{\nu\rangle} + \lambda^{\langle\mu\nu\alpha\rangle}_{\mathbf{k}} k_{\langle\mu}k_\nu k_{\alpha\rangle} + \lambda^{\langle\mu\nu\alpha\beta\rangle}_{\mathbf{k}} k_{\langle\mu}k_\nu k_\alpha k_{\beta\rangle} + \mathcal{O} \left( k^5 \right). \label{eq_30moment}
\end{equation}
We note that, recently, the truncation of the expansion of the distribution function in momenta of rank 3 has also been studied in the context of relativistic shock waves \cite{calzetta}. The expansion coefficients $\lambda_{\mathbf{k}}^{\langle\mu_1\cdots\mu_\ell\rangle}$ are given by, 
\begin{subequations}
\begin{align}
\lambda_{\mathbf{k}}&=\Phi_0+P^{(0)}_{\mathbf{k}1}\Phi_1+P^{(0)}_{\mathbf{k}2}\Phi_2,\\
\lambda_{\mathbf{k}}^{\langle\mu\rangle}&=\Phi^{\langle\mu\rangle}_0+P^{(1)}_{\mathbf{k}1}\Phi^{\langle\mu\rangle}_1,\\
\lambda_{\mathbf{k}}^{\langle\mu\nu\rangle}&=\Phi^{\langle\mu\nu\rangle}_0,\\
\lambda_{\mathbf{k}}^{\langle\mu\nu\alpha\rangle}&=\Phi^{\langle\mu\nu\alpha\rangle}_0,\\
\lambda_{\mathbf{k}}^{\langle\mu\nu\alpha\beta\rangle}&=\Phi^{\langle\mu\nu\alpha\beta\rangle}_0.
\end{align}
\end{subequations}
The expressions for the expansion coefficients of rank 0, 1, and 2 are identical to the ones used in the 14-moment approximation, see Eqs.~\eqref{eq:lambda_DNMR}. The expansion coefficients of rank 3 and 4 are new and guarantee that irreducible moments of rank 3 and 4 can be introduced as novel dynamical variables. This is the minimal truncation scheme for a linearly causal and stable third-order theory. We note that these new terms in the expansion of $\phi_{\mathbf{k}}$ increases the number of independent fields from 14 to 30, since each tensor contributes with $2\ell+1$ degrees of freedom, with $\ell$ being the rank of the respective tensor, given that they are symmetric, traceless and orthogonal to $u^\mu$ in all indices. 

\subsection{Equations of motion}

So far, we have truncated the expansion for the nonequilibrium distribution function, imposing that it can be determined in terms of 30 degrees of freedom. This procedure will naturally lead to a closed set of equations of motion for such variables. For the sake of convenience, we shall now derive relations between the irreducible moments of the nonequilibrium distribution function and the coefficients of the truncated moment expansion. In other words, our goal is to obtain relations between $\Phi^{\mu_1\cdots\mu_\ell}_r$ and the irreducible moments $\rho^{\mu_1\cdots\mu_\ell}_r$. From Eqs.~\eqref{phi_series}-\eqref{def_P}, one can show that
\begin{equation}
\rho^{\mu_1\cdots\mu_\ell}_r = (-1)^\ell \, \ell! \sum_{n=0}^{N_\ell} \sum_{m=0}^{n} \Phi_n^{\langle\mu_1\cdots\mu_\ell\rangle} a^{(\ell)}_{nm} J_{r+m+2\ell,\ell}.
\end{equation}
Therefore, given the truncation scheme adopted ($N_0 = 2$, $N_1 = 1$, and $N_2 = N_3 = N_4 = 0$, with higher-rank contributions being completed neglected), it follows that the irreducible moments are given by 
\begin{subequations}
\label{moments_currents}
\begin{align}
\rho_r&=\gamma_r^\Pi \Pi,\\
\rho^\mu_r&=\gamma^n_r n^\mu,\\
\rho^{\mu\nu}_r&=\gamma^\pi_r \pi^{\mu\nu},\\
\rho^{\mu\nu\alpha}_r&=\gamma^\Omega_r \Omega^{\mu\nu\alpha}, \label{eom3_Phi}  \\
\rho^{\mu\nu\alpha\beta}_r&=\gamma^\Theta_r \Theta^{\mu\nu\alpha\beta}, \label{eom4_Phi}
\end{align}
\end{subequations}
where we have defined $\Omega^{\mu\nu\alpha} \equiv \rho_0^{\mu\nu\alpha}$ and $\Theta^{\mu\nu\alpha\beta} \equiv \rho_0^{\mu\nu\alpha\beta}$, and introduced the following thermodynamic coefficients
\begin{subequations}
\begin{align}
\gamma^\Pi_r &= \mathcal{A}_\Pi J_{r,0}+\mathcal{A}_\Pi J_{r+1,0}+\mathcal{C}_\Pi J_{r+2,0}, \\
\gamma^n_r &= -\frac{J_{41} J_{r+2,1}}{D_{31}} + \frac{J_{31}J_{r+3,1}}{D_{31}} , \\
\gamma^\pi_r &= \frac{J_{r+4,2}}{J_{4,2}}, \,
\gamma^\Omega_r = \frac{J_{r+6,3}}{J_{6,3}}, \,
\gamma^\Theta_r = \frac{J_{r+8,4}}{J_{8,4}},
\end{align}
\end{subequations}
with
\begin{subequations}
\begin{align}
\mathcal{A}_\Pi&=-\frac{3}{m^2}\frac{D_{30}}{J_{20}D_{20}+J_{30}G_{12}+J_{40}D_{10}},\\
\mathcal{B}_\Pi&=-\frac{3}{m^2}\frac{G_{23}}{J_{20}D_{20}+J_{30}G_{12}+J_{40}D_{10}},\\
\mathcal{C}_\Pi&=-\frac{3}{m^2}\frac{D_{20}}{J_{20}D_{20}+J_{30}G_{12}+J_{40}D_{10}}.
\end{align}
\end{subequations}

In particular, note that there is an infinite number of equations of motion for the irreducible moments, labeled by the subindices $r$, cf.~Eqs.~\eqref{eom0}-\eqref{eom4}. Therefore, there is an ambiguity in the choice of the dynamic variable of the theory: one has the freedom to take any particular value for $r$ and construct the theory around the corresponding irreducible moment \cite{Denicol:2010xn}. Following Ref.~\cite{Denicol:2010xn}, we choose $r=0$, hence taking the equations of motion for the irreducible moments $\rho^{\mu_1\cdots\mu_\ell}_0$ as the starting point for our derivation.

A relation for the generalized collision term in terms of the irreducible moments is still required. For the sake of simplicity, in this work we employ the relaxation time approximation \cite{oldRTA, nRTA}. For the matching conditions employed in this work and assuming an energy-independent relaxation time, this approximation yields
\begin{equation}
C[f] = - \frac{E_\mathbf{k}}{t_R} \delta f_{\mathbf{k}},
\end{equation}
with $t_R$ being the relaxation time. In this case, we obtain a rather simple expression for the generalized collision term, given by
\begin{equation}
C^{\langle \mu_1 \cdots \mu_\ell \rangle}_{r - 1} = - \frac{\rho^{\mu_1 \cdots \mu_\ell}_r}{t_R}. \label{eqRTA}
\end{equation}
We remark that such prescription for the collision term is only consistent with the conservation of energy and momentum as long as one imposes Landau matching conditions. A generalization of the relaxation time approximation for which the conservation laws are fulfilled for arbitrary matching conditions was first addressed in Ref.~\cite{nRTA}.

We are now in position to obtain a closed set of equations of motion for the dissipative currents, $\Pi$, $n^\mu$ and $\pi^{\mu\nu}$, and, more importantly, for the novel fields $\Omega^{\mu\nu\lambda}$ and $\Theta^{\mu\nu\alpha\beta}$. For this purpose, we insert the relations in Eqs.~\eqref{moments_currents} and \eqref{eqRTA} into the equations of motion for the irreducible moments, Eqs.~\eqref{eom0}-\eqref{eom4}, for $r=0$. 
We then obtain, for the bulk viscous pressure,
\begin{equation}
\tau_\Pi\dot{\Pi}+\Pi=-\zeta\theta-\tau_{\Pi n}n_\mu \nabla^\mu P_0-\ell_{\Pi n}\partial_\mu n^\mu-\delta_{\Pi\Pi}\Pi\theta+\lambda_{\Pi n}n^\mu\nabla_\mu\alpha_0+\lambda_{\Pi\pi}\pi^{\mu\nu}\sigma_{\mu\nu}. \label{eom_bulk}
\end{equation}
For the particle diffusion, we have
\begin{eqnarray}
\tau_n\dot{n}^{\langle\mu\rangle}+n^\mu&=& \kappa_n \nabla^\mu \alpha_0- \tau_n n_\nu \omega^{\nu\mu}-\delta_{nn}n^\mu\theta+\tau_{n\Pi}\Pi\nabla^\mu P_0-\tau_{n\pi}\pi^{\mu\nu}\nabla_\nu P_0-\ell_{n\Pi}\nabla^\mu\Pi+ \notag \\
&+&\ell_{n\pi}\Delta^\mu_\alpha\partial_\beta\pi^{\alpha\beta}-\lambda_{nn}n_\nu\sigma^{\mu\nu}+\lambda_{n\Pi}\Pi\nabla^\mu\alpha_0-\lambda_{n\pi}\pi^{\mu\nu}\nabla_\nu\alpha_0-\tau_n\gamma^\Omega_{-2}\Omega^{\mu\alpha\beta}\sigma_{\alpha\beta}. \label{eom_diffusion}
\end{eqnarray}
The equation of motion for the shear-stress tensor reads
\begin{eqnarray}
\tau_\pi\dot{\pi}^{\langle\mu\nu\rangle}+\pi^{\mu\nu} &=& 2\eta\sigma^{\mu\nu} + 2 \tau_\pi \pi^{\langle\mu}_\lambda\omega^{\nu\rangle\lambda}-\delta_{\pi\pi}\pi^{\mu\nu}\theta-\tau_{\pi\pi}\pi^{\langle\mu}_\lambda \sigma^{\nu\rangle\lambda}+\lambda_{\pi\Pi}\Pi\sigma^{\mu\nu}-\tau_{\pi n}n^{\langle\mu}\nabla^{\nu\rangle}P_0+\ell_{\pi n}\nabla^{\langle\mu}n^{\nu\rangle}+\notag \\
&+&\lambda_{\pi n}n^{\langle\mu}\nabla^{\nu\rangle}\alpha_0+\tau_\pi\left(-\gamma^\Omega_{-1}\Delta^{\mu\nu}_{\alpha\beta}\nabla_\lambda\Omega^{\alpha\beta\lambda}+\lambda_{\pi\Omega}\Omega^{\mu\nu\lambda}\nabla_\lambda\alpha_0+\tau_{\pi\Omega}\dot{u}_\alpha\Omega^{\mu\nu\alpha}-\gamma_{-2}^\Theta\Theta^{\mu\nu\alpha\beta}\sigma_{\alpha\beta}\right). \label{eom_shear}
\end{eqnarray}
We remark that these equations are identical to the ones obtained in Ref.~\cite{dnmr}, with the exception of the third-order corrections that are considered in the present work. The expressions for the novel transport coefficients of the theory are explicitly listed in Appendix \ref{app_coeffs}. The coefficients that already appeared in the second-order version of this formulation can be found in Ref.~\cite{dnmr}. 

The dissipative currents are now coupled to novel degrees of freedom, which satisfy their own equations of motion. First, the equation of motion for $\Omega^{\mu \nu \alpha}$ is derived substituting Eq.~\eqref{eom3_Phi} into Eq.~\eqref{eom3} for $r=0$, leading to
\begin{eqnarray}
\tau_\Omega \dot{\Omega}^{\langle\mu\nu\alpha\rangle} + \Omega^{\mu\nu\alpha}
& = & 
\delta_{\Omega \Omega} \Omega^{\mu\nu\alpha}\theta + \ell_{\Omega n} \sigma^{\langle\mu\nu} n^{\alpha\rangle} + 3 \tau_\Omega \Omega^{\lambda\langle\mu\nu}\omega^{\alpha\rangle}{}_\lambda + \tau_{\Omega \Omega} \sigma^{\langle\mu}_\lambda \Omega^{\nu\alpha\rangle\lambda} + \frac{3}{7} \eta_\Omega \nabla^{\langle\mu} \pi^{\nu\alpha\rangle} + \lambda_{\Omega\pi} \pi^{\langle\mu\nu} \nabla^{\alpha\rangle} \alpha \notag \\
& + &
\tau_{\Omega \pi}\pi^{\langle\mu\nu} \nabla^{\alpha\rangle} P - 3 \tau_\Omega \gamma_1^\pi \pi^{\langle\mu\nu}\dot{u}^{\alpha\rangle} + \lambda_{\Omega\Theta} \Theta^{\mu\nu\alpha\beta} \nabla_\beta \alpha + \tau_{\Omega\Theta} \Theta^{\mu\nu\alpha\beta} \dot{u}_\beta - \tau_\Omega \gamma_{-1}^\Theta \Delta^{\mu\nu\alpha}_{\lambda\sigma\rho}\nabla_\beta \Theta^{\lambda\sigma\rho\beta}. \label{eom_rank3}
\end{eqnarray}
Finally, substituting Eq.~\eqref{eom4_Phi} into Eq.~\eqref{eom4} for $r=0$, one obtains an equation of motion for
$\Theta^{\mu\nu\alpha\beta}$,
\begin{eqnarray}
\tau_\Theta\dot{\Theta}^{\langle\mu\nu\alpha\beta\rangle}+\Theta^{\mu\nu\alpha\beta}&=&\delta_{\Theta\Theta}\Theta^{\mu\nu\alpha\beta}\theta+4 \tau_\Theta \Theta^{\lambda\langle\mu\nu\alpha}\omega^{\beta\rangle}{}_\lambda+\tau_{\Theta\Theta}\sigma^{\langle\mu}_\lambda\Theta^{\nu\alpha\beta\rangle\lambda}+\ell_{\Theta\pi}\sigma^{\langle\mu\nu}\pi^{\alpha\beta\rangle}+ \notag \\
&+&\ell_{\Theta\Omega}\nabla^{\langle\mu}\Omega^{\nu\alpha\beta\rangle}+\tau_{\Theta\Omega}\dot{u}^{\langle\mu}\Omega^{\nu\alpha\beta\rangle}+\lambda_{\Theta\Omega}\Omega^{\langle\mu\nu\alpha}\nabla^{\beta\rangle}\alpha_0,
\label{eom_Theta}
\end{eqnarray}
where all transport coefficients in these equations are listed in Appendix \ref{app_coeffs}. We emphasize that, since we employ the relaxation time approximation, all relaxation times are identical $\tau_\Pi = \tau_n = \tau_\pi = \tau_\Omega = \tau_\Theta = t_R$.

These are the third-order fluid-dynamical equations from the method of moments. In the derivation of Eq.~\eqref{eom_Theta}, we have used Eq.~\eqref{eq:udot} to express gradients of the thermodynamic pressure in terms of the time derivative of the fluid 4-velocity, further omitting fourth-order terms. On top of that, we have also used the covariant version of the Gibbs-Duhem equation,
\begin{equation}
\nabla_\mu \beta_0=\frac{1}{\varepsilon_0+P_0}\left(n_0\nabla_\mu\alpha_0-\beta_0\nabla_\mu P_0\right).
\end{equation}

Moreover, causality and stability of a linearized version of this theory were first investigated in Ref.~\cite{bd3order} and such properties are simultaneously fulfilled as long as the transport coefficients satisfy the following inequalities
\begin{eqnarray}
\left[3\tau_\pi\left(1-c_{\mathrm{s}}^2\right)-4\frac{\eta}{\varepsilon_0+P_0}\right]\tau_\Omega & > & \frac{27}{35} \eta_\Omega \, \gamma_{-1}^\Omega \tau_\pi\left(1-c_{\mathrm{s}}^2\right), \label{eq:stability1} \\
3(1-\mathrm{c}_\mathrm{s}^2)\tau_\pi&\geq&\frac{4\eta}{\varepsilon_0 +P_0},
\label{eq:stability2}
\end{eqnarray}
which are Eqs.~(104) and (105) of the above-mentioned paper adapted to the notation used in the present work ($\eta_\rho \rightarrow \eta_\Omega \, \gamma_{-1}^\Omega$ and $\tau_\rho \rightarrow \tau_\Omega$). The second condition is well known and is also applicable to second-order theories while the first condition is specific to third-order theories.
In particular, with all the assumptions considered in this work, i.e., the classical and massless limits, as well as the relaxation time approximation, the conditions above simplify, respectively, to
\begin{eqnarray}
\eta_\Omega & < & \frac{49}{3} T \, \tau_\pi, \\
\tau_\pi & \geq & \frac{2 \eta}{\varepsilon_0 +P_0},
\end{eqnarray}
where we have used that $\gamma_{-1}^\Omega = 1/(7 \, T)$ in the classical and massless limits. Our results for the transport coefficients, $\tau_\pi = 5\eta/(\varepsilon_0 + P_0)$ and $\eta_\Omega = 6 \, T \, \tau_\pi$, listed in Appendix \ref{app_coeffs}, are thus consistent with the fundamental conditions listed above.

\section{Bjorken flow}
\label{sec_bjorken}

We are interested in analyzing the agreement between the solutions of the third-order equations of motion with solutions of the relativistic Boltzmann equation. A particularly convenient configuration to perform this study is the Bjorken flow \cite{bjorken}, a highly symmetric framework constructed as a toy model for studying relativistic heavy-ion collisions. In this case, it is rather convenient to employ Milne coordinates, which are related to Cartesian coordinates through 
\begin{equation}
\tau\equiv\sqrt{t^2-z^2}, \hspace{.3cm} \eta_s\equiv\tanh^{-1}\left(\frac{z}{t}\right),
\end{equation}
with $\tau$ being the proper time and $\eta_s$ being the spacetime rapidity. The first is invariant under Lorentz-boosts in the $z$-direction, while the second is simply shifted by a constant value under such boosts. Milne coordinates are described by the following metric tensor
\begin{equation}
g_{\mu\nu}=\text{diag}(1,-1,-1,-\tau^2),
\end{equation}
and therefore the only nonvanishing Christoffel symbols are
\begin{equation}
\Gamma^\tau_{\eta_s \eta_s}=\tau,\hspace{.3cm}\Gamma^{\eta_s}_{\tau\eta_s}=\Gamma^{\eta_s}_{\eta_s\tau}=\frac{1}{\tau}.
\end{equation}

A series of assumptions is taken for simplicity. First, it is assumed that the fluid is homogeneous in this coordinate system. That is, all the fluid-dynamical fields do not depend on the spacetime coordinates, $x$, $y$ and $\eta_s$, only on the proper time, $\tau$. We further assume that the system is symmetric under rotations in the transverse $xy$--plane and is symmetric under reflections around the longitudinal $\eta_s$-axis. 

In Milne coordinates, we assume the trivial solution for the fluid 4-velocity,  $u^\mu=(u^\tau,u^x,u^y,u^{\eta_s})=(1,0,0,0)$. We also note that the aforementioned assumptions further imply that all 4-vectors that are orthogonal to $u^\mu$ vanish, since there cannot be any preferred direction in the transverse $xy$-plane and in the longitudinal $\eta_s$-axis. Therefore, in Bjorken flow, both the particle diffusion 4-current, $n^\mu$, and the energy diffusion 4-current, $W^\mu$, are zero. Moreover, it can be shown that spatial gradients of scalar functions are identically zero \cite{Denicol:2021}. The shear tensor and shear-stress tensor, on the other hand, can be expressed in the following form
\begin{equation}
\sigma_{\mu\nu}=\text{diag}\left(0,\frac{1}{3\tau},\frac{1}{3\tau},-\frac{2\tau}{3}\right),\hspace{.4cm}\pi^{\mu\nu}=\text{diag}\left(0,\frac{\pi}{2},\frac{\pi}{2},-\frac{\pi}{\tau^2}\right),
\end{equation}
while the expansion rate is given by $\theta=1/\tau$. Therefore, the conservation of energy, Eq.~\eqref{energy_cons}, reduces to
\begin{equation}
\frac{d\varepsilon}{d\tau}=-\frac{1}{\tau}\left(\varepsilon+P-\pi\right). \label{edo_energy_bjorken}
\end{equation}
The momentum conservation equation, Eq.~\eqref{momentum_cons}, is trivially satisfied. In order to compare our results to those of Ref.~\cite{amaresh}, we neglect any contribution of the particle density, setting it to zero, and Eq.~\eqref{charge_cons} becomes trivially satisfied as well.

Since we consider a classical gas of massless particles, the energy density and thermodynamic pressure are related through $\varepsilon=3P$. On top of that, the energy density is a quartic function of the temperature, $\varepsilon\sim T^4$. It is then convenient to rewrite Eq.~\eqref{edo_energy_bjorken} as a differential equation for the temperature,
\begin{equation}
\frac{dT}{d\tau}=\frac{T}{3\tau}(\hat{\pi}-1), \label{eq:dT/dtau}
\end{equation} 
with $\hat{\pi} \equiv \pi/(\varepsilon_0+P_0)$.

The next step is to obtain the equations of motion for the dissipative currents that couple with the conservation of energy and momentum in Bjorken flow. First, we note that the bulk viscous pressure is zero, since we are considering a system of massless particles, cf.~Eq.~\eqref{quantities_rho}. As already stated, the particle diffusion 4-current is identically zero in Bjorken flow, since it is orthogonal to the 4-velocity. Furthermore, all irreducible moments of odd rank also vanish in this framework \cite{Denicol:2021}. Wherefore, the relevant fluid-dynamical equations reduce to
\begin{subequations}
\begin{align}
\tau_\pi\dot{\pi}^{\langle\mu\nu\rangle}+\pi^{\mu\nu}&=2\eta\sigma^{\mu\nu}-\delta_{\pi\pi}\pi^{\mu\nu}\theta-\tau_{\pi\pi}\pi^{\langle\mu}_\lambda \sigma^{\nu\rangle\lambda}-\tau_\pi\gamma_{-2}^\Theta\Theta^{\mu\nu\alpha\beta}\sigma_{\alpha\beta}, \label{eom_shear_bjorken} \\
\tau_\Theta\dot{\Theta}^{\langle\mu\nu\alpha\beta\rangle}+\Theta^{\mu\nu\alpha\beta}&=\delta_{\Theta\Theta}\Theta^{\mu\nu\alpha\beta}\theta+\tau_{\Theta\Theta}\sigma^{\langle\mu}_\lambda\Theta^{\nu\alpha\beta\rangle\lambda}+\ell_{\Theta\pi}\sigma^{\langle\mu\nu}\pi^{\alpha\beta\rangle}. \label{eom_theta_bjorken}
\end{align}
\end{subequations}
In the massless limit, these transport coefficients are
\begin{equation}
\tau_\pi=\frac{5\eta}{Ts}, \hspace{.2cm} \frac{\delta_{\pi\pi}}{\tau_\pi}=\frac{4}{3}, \hspace{.2cm} \frac{\tau_{\pi\pi}}{\tau_\pi}=\frac{10}{7}, \hspace{.2cm} \gamma_{-2}^\Theta=\frac{1}{72T^2}, \hspace{.2cm} \frac{\delta_{\Theta\Theta}}{\tau_\Theta}=-2, \hspace{.2cm}\frac{\tau_{\Theta\Theta}}{\tau_\Theta}=-\frac{36}{11}, \hspace{.2cm} \ell_{\Theta\pi}=64T^2,
\end{equation}
where $\eta$ is the shear viscosity coefficient and $s$ is the entropy density. We remark that the first three transport coefficients were first calculated in Ref.~\cite{dnmr}, while the last four were calculated in this work. Once again, general expressions for the latter can be found in Appendix \ref{app_coeffs}.

It is convenient to define a unitary 4-vector, $z_\mu = (0,0,0,1)$, and project Eqs.~\eqref{eom_shear_bjorken} and \eqref{eom_theta_bjorken} with $z_\mu z_\nu$ and $z_\mu z_\nu z_\alpha z_\beta$, respectively, in order to obtain a closed equation of motion for the longitudinal components of $\pi^{\mu\nu}$ and $\Theta^{\mu\nu\alpha\beta}$. These equations then become
\begin{subequations}
\begin{align}
\frac{d\hat{\pi}}{d\tau}&=-\frac{\hat{\pi}}{\tau_\pi}+\frac{4}{15\tau}-\frac{10}{21}\frac{\hat{\pi}}{\tau}-\frac{4}{3}\frac{\hat{\pi}^2}{\tau}-\frac{1}{72}\frac{\hat{\varphi}}{\tau}, \label{eom_scalar_shear_bjorken} \\
\frac{d\hat{\varphi}}{d\tau}&=-\frac{\hat{\varphi}}{\tau_\Theta}+\frac{768}{35}\frac{\hat{\pi}}{\tau} -\frac{60}{77}\frac{\hat{\varphi}}{\tau}-2\frac{\hat{\varphi} \hat{\pi}}{\tau}, \label{eq:eom_varphi}
\end{align}
\end{subequations}
where we have used Eq.~\eqref{eq:dT/dtau} to obtain an equation of motion for the dimensionless variable $\hat{\varphi}\equiv\Theta^{\eta_s \eta_s}_{\eta_s\eta_s}/[(\varepsilon+P)T^2]$ and employed the transport coefficients given in Ref.~\cite{dnmr}.

\begin{figure}[ht]
\begin{center}
\includegraphics[width=.48\textwidth]{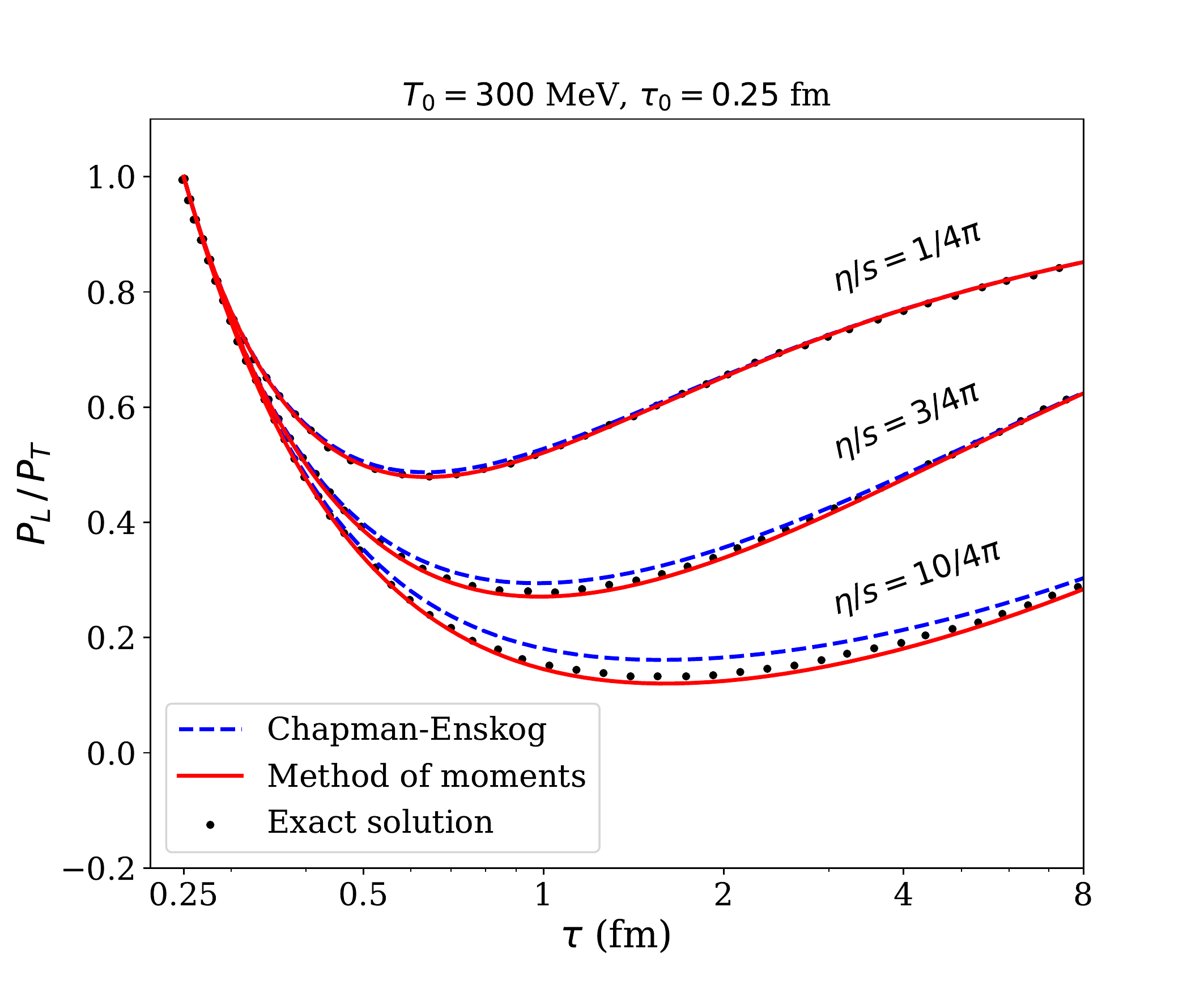}
\includegraphics[width=.48\textwidth]{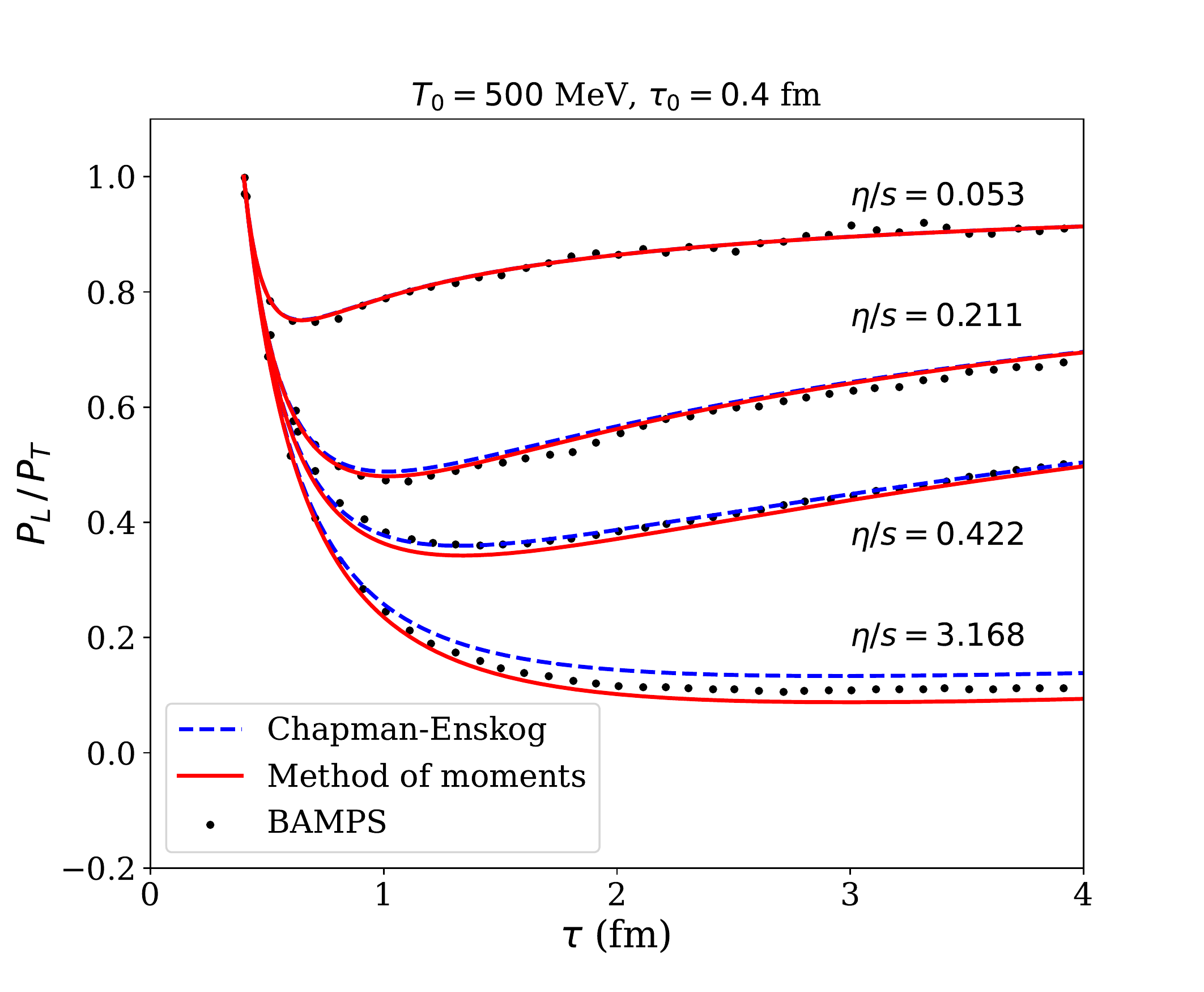}
\caption{(Color online) Pressure anisotropy in Bjorken flow for RHIC (left panel) and LHC (right panel) energies, as calculated from the Chapman-Enskog method \cite{amaresh}, method of moments and solutions of the Boltzmann equation from BAMPS for several values of $\eta/s$, considering $\tau_\Theta=\tau_\pi=\frac{5\eta}{\varepsilon+P}$ \cite{dnmr}.}
\label{fig:pressure_aniso}
\end{center}
\end{figure}

In Fig.~\ref{fig:pressure_aniso}, we compare the results for the pressure anisotropy in Bjorken flow, defined as $P_L/P_T = (1-4\hat{\pi})/(1+2\hat{\pi})$, calculated within the third-order formalism developed in Ref.~\cite{amaresh} (blue dashed lines), the one proposed in this work (red solid lines) and solutions of the Boltzmann equation (black dots). In the left panel, we compare to solutions of the Boltzmann equation calculated assuming the relaxation time approximation, with initial time and temperature calibrated to describe the matter produced in heavy-ion collisions at RHIC energies \cite{Florkowski_2013}. In the right panel, we compare to solutions of the Boltzmann equation calculated using the Boltzmann Approach To Multi-Parton Scatterings (BAMPS) \cite{El_2010}. The initial time and temperature were calibrated to describe the matter produced in heavy-ion collisions at LHC energies. In both scenarios, we have assumed an initially isotropic pressure configuration, $P_L/P_T = 1$. We see that solutions of both third-order fluid-dynamical theories are in good agreement with solutions of the microscopic theory, with the method of moments displaying a slightly better description. We remark that BAMPS solves the full Boltzmann equation without relying on the relaxation time approximation for the collision term. Thus, the agreement with the fluid-dynamical calculations may suggest that this approximation is reasonable, at least for the purposes of describing the time evolution of the shear-stress tensor. 

\begin{figure}[ht]
\begin{center}
\includegraphics[width=.48\textwidth]{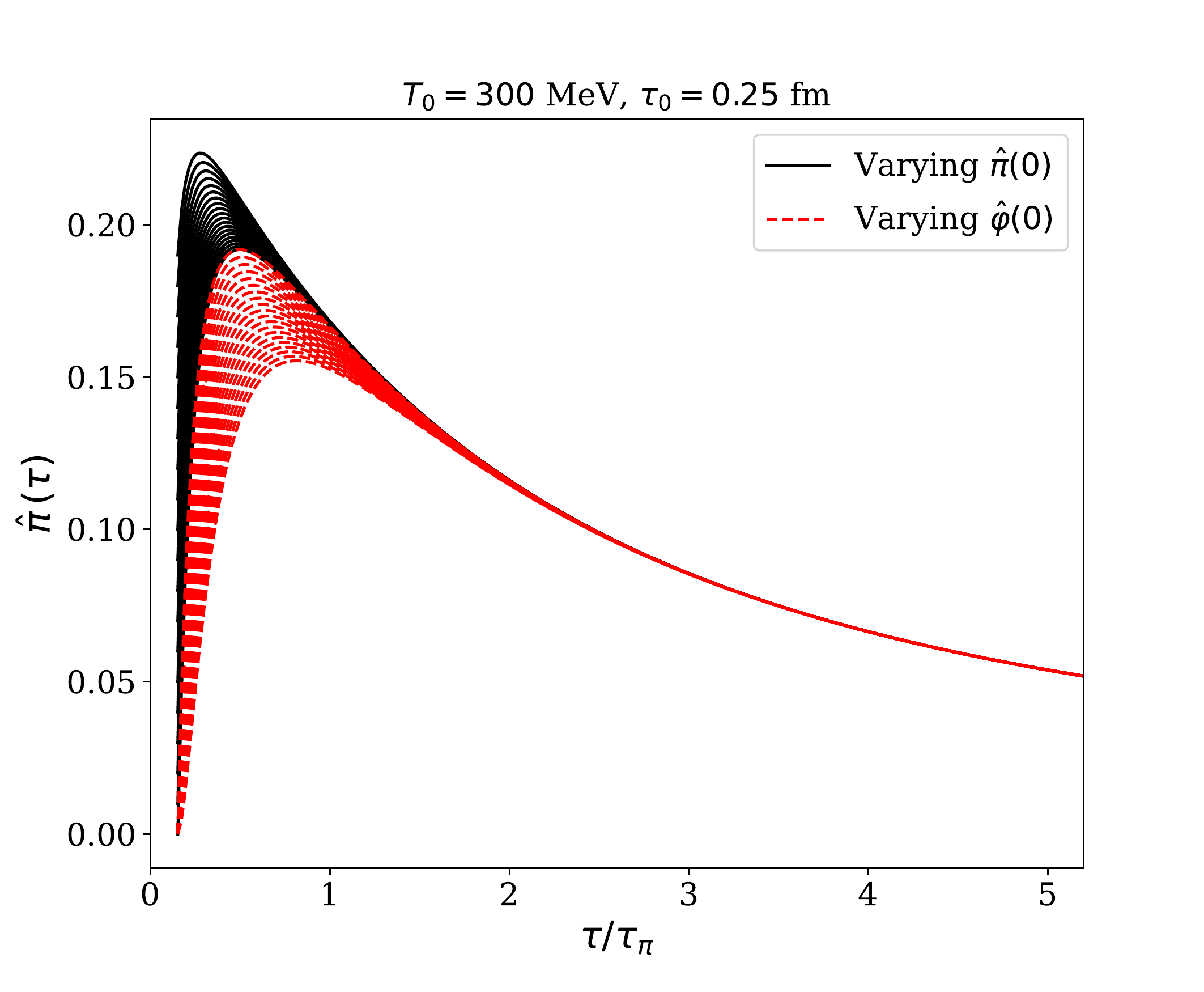}
\includegraphics[width=.48\textwidth]{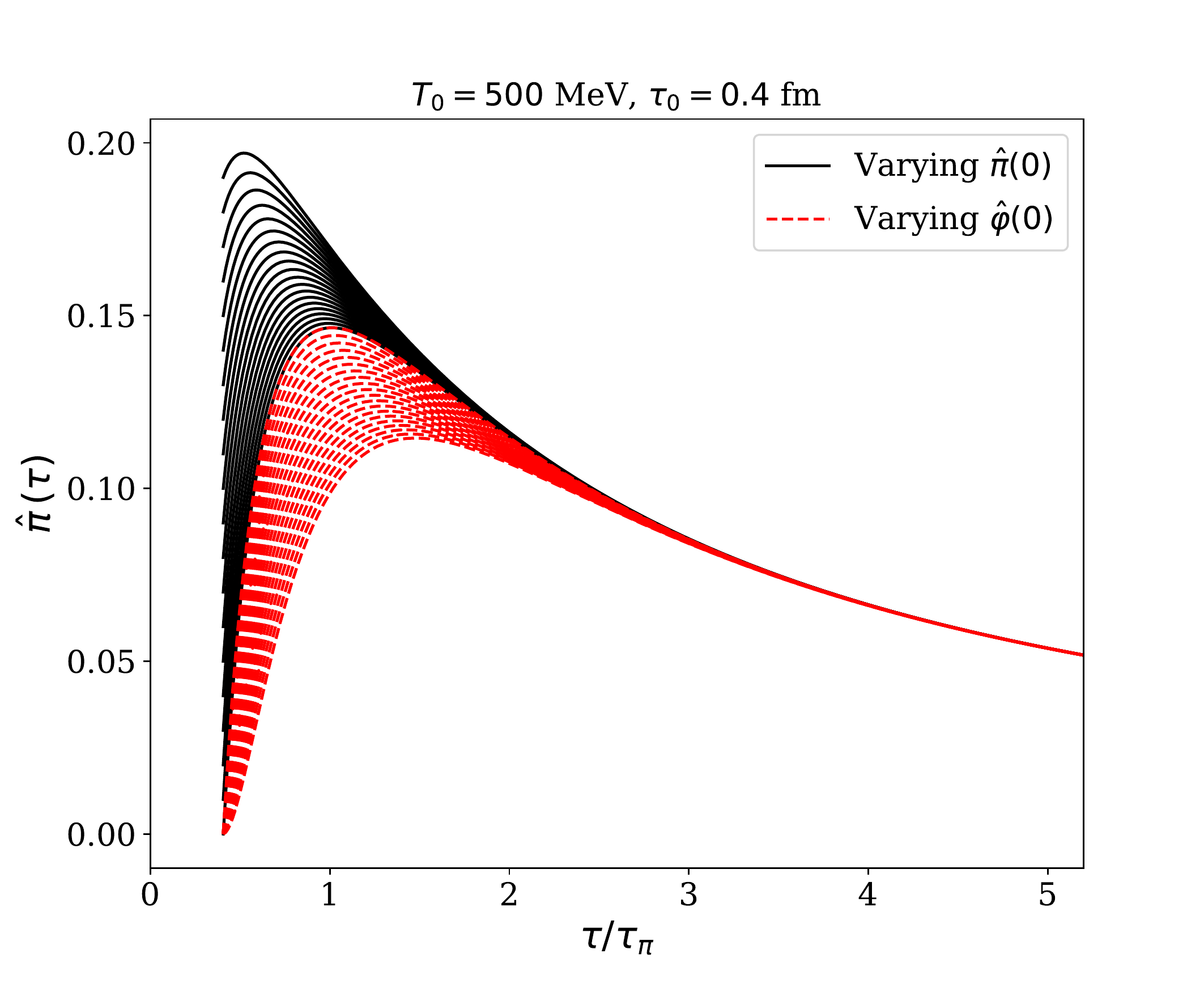}
\caption{(Color online) $\hat{\pi}$ as a function of $\tau/\tau_\pi$ for RHIC (left panel) and LHC (right panel) energies, considering several initial conditions for $\hat{\pi}$ and $\hat{\varphi}$, assuming $\tau_\Theta=\tau_\pi=\frac{5\eta}{\varepsilon+P}$ \cite{dnmr}.}
\label{fig:shear}
\end{center}
\end{figure}

\begin{figure}[ht]
\begin{center}
\includegraphics[width=.48\textwidth]{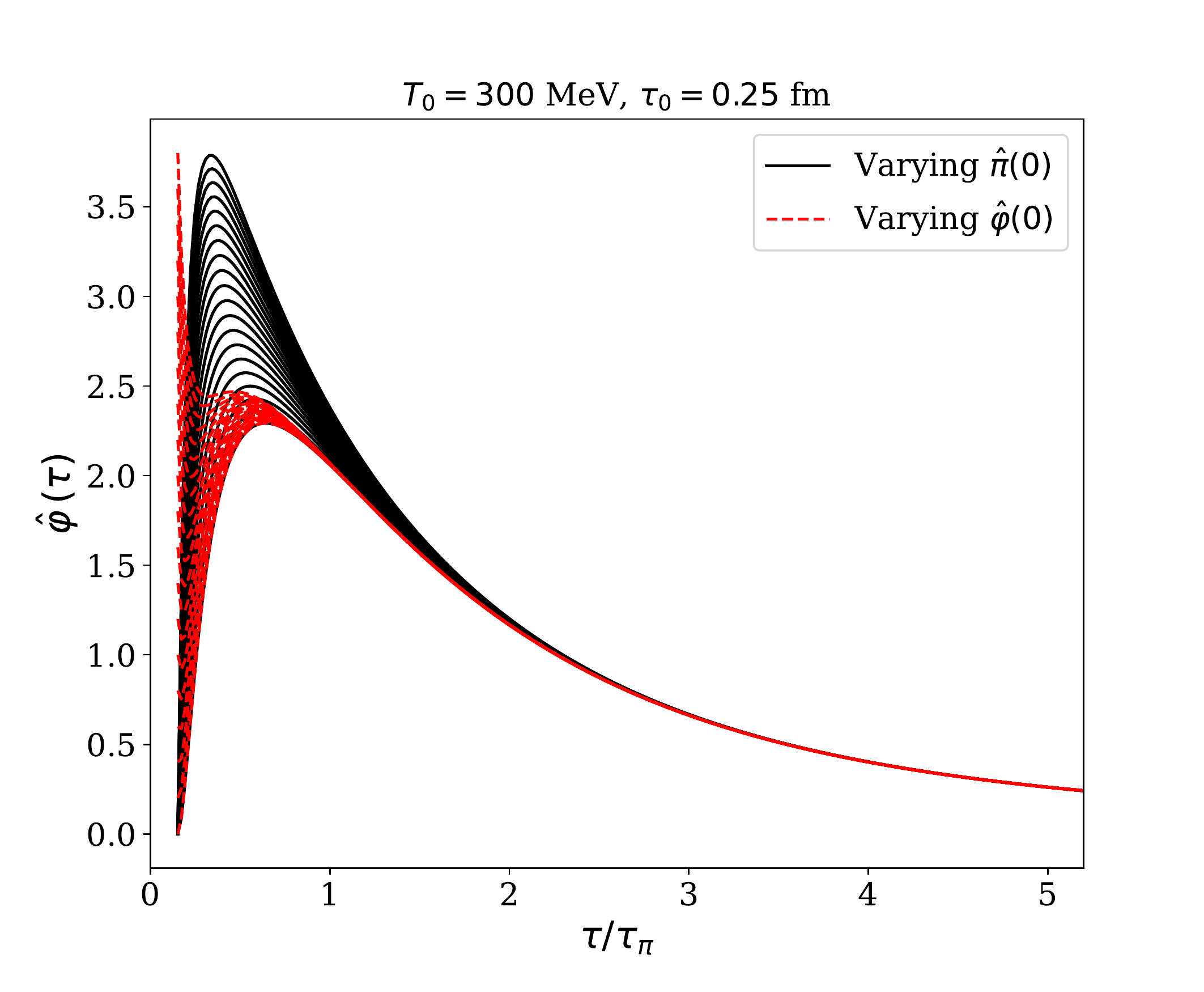}
\includegraphics[width=.48\textwidth]{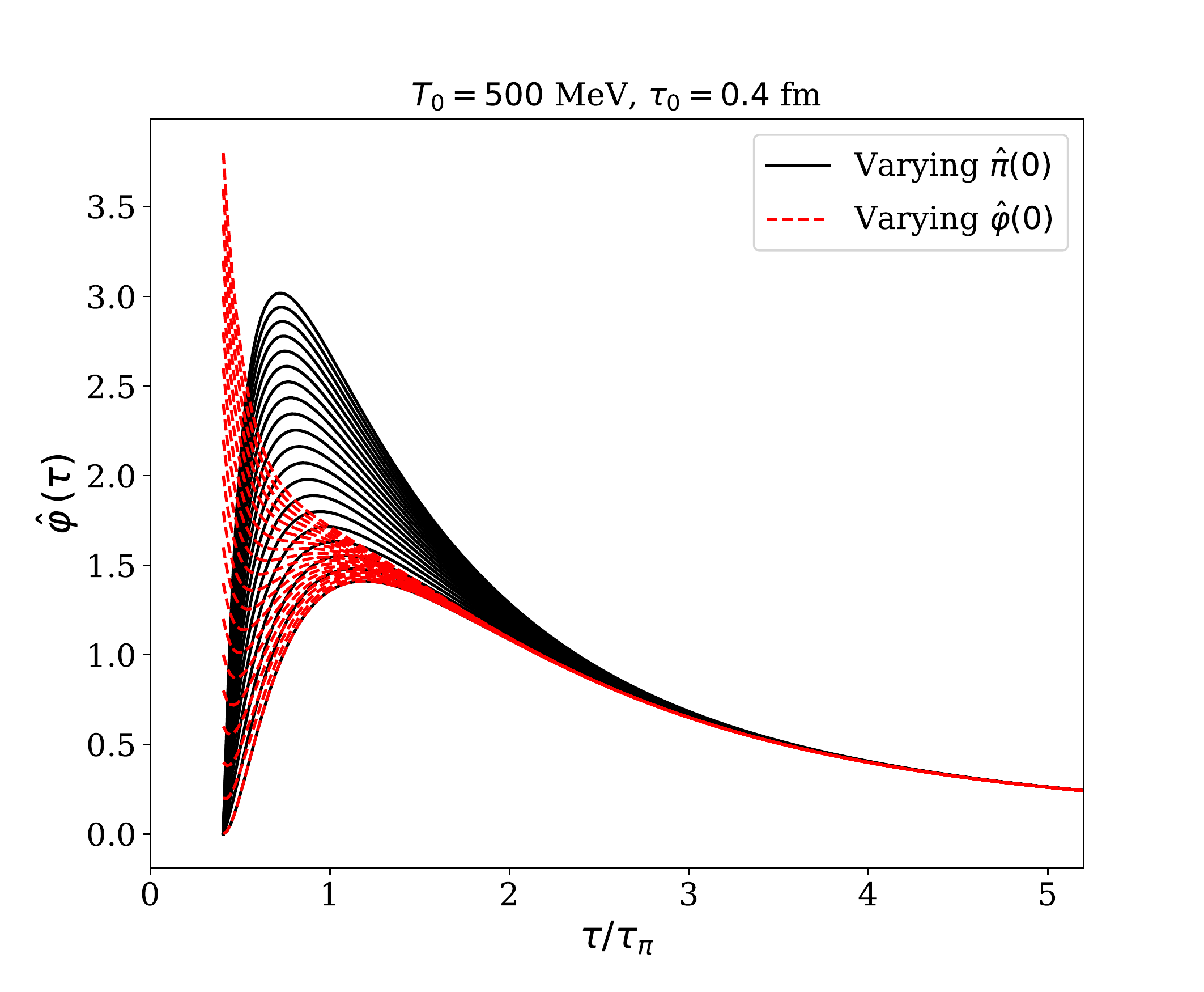}
\caption{(Color online) $\hat{\varphi}$ as a function of $\tau/\tau_\pi$ for RHIC (left panel) and LHC (right panel) energies, considering several initial conditions for $\hat{\pi}$ and $\hat{\varphi}$, assuming $\tau_\Theta=\tau_\pi=\frac{5\eta}{\varepsilon+P}$ \cite{dnmr}.}
\label{fig:rank4}
\end{center}
\end{figure}

For the sake of completeness, in Figs.~\ref{fig:shear} and \ref{fig:rank4}, we display $\hat{\pi}$ and $\hat{\varphi}$, respectively, as function of $\tau/\tau_\pi$, for a wide set of initial values of $\hat{\pi}$ (black solid lines) and $\hat\varphi$ (red dashed lines), considering both the RHIC and LHC scenarios described above. In both cases, we observe that these quantities approach the same universal values at large proper times, regardless of which set of initial conditions is being used. This universal behavior displayed by the fluid-dynamical variables at late times in spite of the initial conditions is called the \textit{hydrodynamic attractor} and was first investigated in Ref.~\cite{heller}. Here, we see that the novel field $\hat\varphi$ also displays this attractor behavior.

\begin{figure}[ht]
\begin{center}
\includegraphics[width=.48\textwidth]{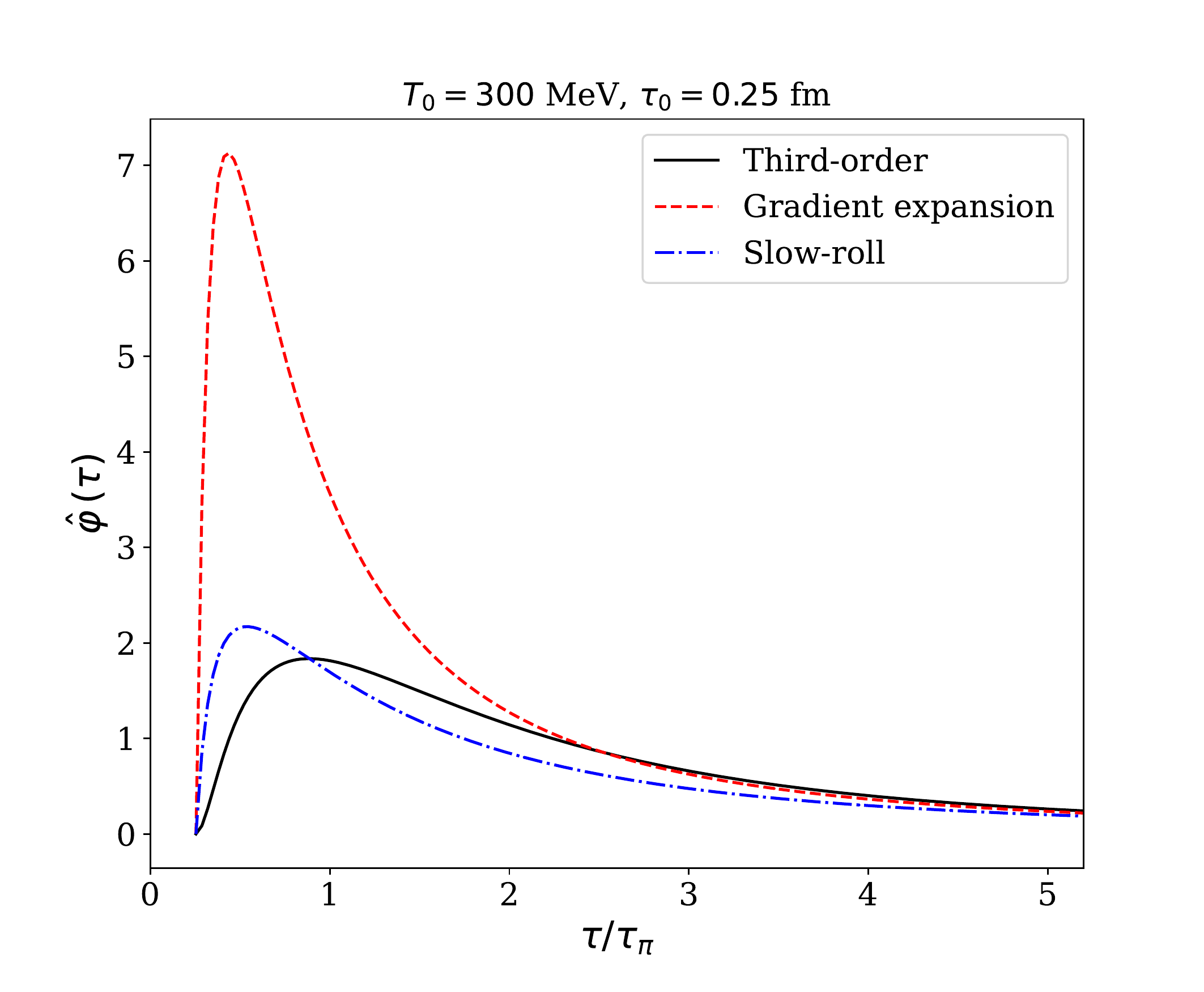}
\includegraphics[width=.48\textwidth]{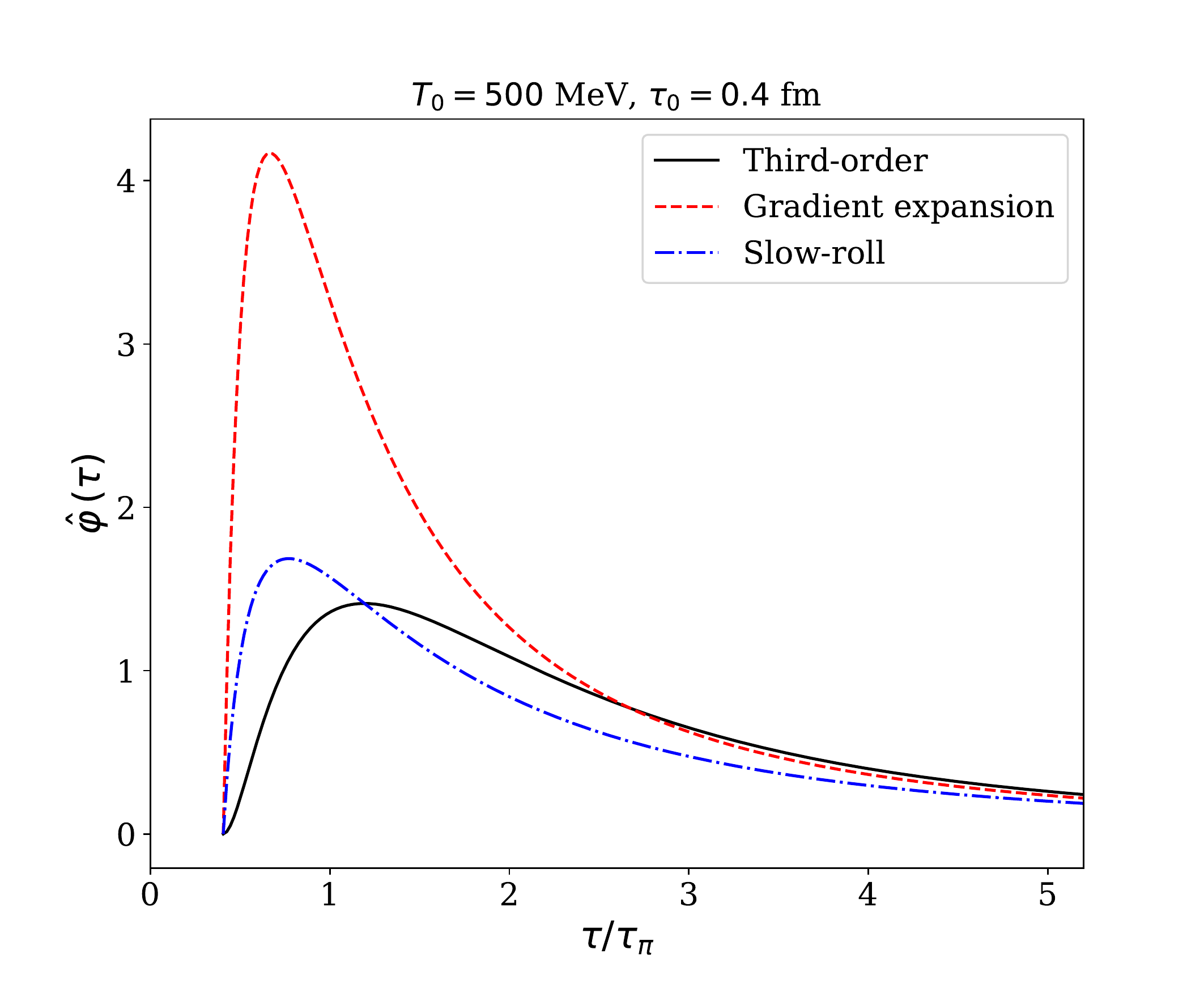}
\caption{(Color online) Solution of $\hat{\varphi}$, starting from equilibrium, as a function of $\tau/\tau_\pi$ and for $\eta/s = 0.5$, compared to its respective Navier-Stokes limit and zeroth-order slow-roll solution. Left panel shows the comparison for RHIC energies while the right panel shows the comparison for LHC energies.}
\label{fig:comparison}
\end{center}
\end{figure}

Last, in Fig.~\ref{fig:comparison}, we compare a solution of Eq.~\eqref{eq:eom_varphi} to two of its asymptotic solutions: (i) its lowest contribution in a gradient expansion, $\hat{\varphi}_{\mathrm{grad}} = 768\hat{\pi}/(35\tau)$ and (ii) its zeroth order slow-roll solution \cite{Liddle_1994, heller, denicol2018analytical, Denicol:2018pak}, obtained by setting $\dot{\hat{\varphi}}=0$, i.e.,
\begin{equation}
\hat\varphi_{\mathrm{slow-roll}} = \frac{768 \hat{\pi}}{35 \left( \frac{\tau}{\tau_\Theta} + \frac{60}{77} + 2 \hat{\pi} \right)}. 
\end{equation}
We consider LHC and RHIC energies and systems that are initially in local equilibrium. In both cases, we observe that the lowest order gradient expansion value of $\hat\varphi$ can surpass its third-order solution by a factor of $\sim 4$, while the zeroth slow-roll solution provides a considerably better agreement with the actual solution at early times. On the other hand, the gradient expansion leading solution converges to the hydrodynamic attractor faster than the slow-roll solution. In the end, both asymptotic values do not provide a good description for the time evolution of $\hat\varphi$.

\section{Conclusions}
\label{sec_conc}

In this work, we have formally derived a linearly causal and stable third-order fluid-dynamical theory from the Boltzmann equation using the method of moments. We demonstrated that equations of motion that include all terms that are asymptotically of third order in gradients can only be obtained including novel degrees of freedom, corresponding to irreducible tensors of rank $3$ and $4$. This is in contrast to the fluid-dynamical theories developed so far, the so-called second-order theories, that only require the inclusion of irreducible tensors of rank 0, 1, and 2 -- which are usually matched to the traditional fluid-dynamical variables appearing in the conserved currents. We generalized the minimal truncation scheme derived by Israel and Stewart \cite{is1}, so that these novel degrees of freedom are taken into account in the derivation procedure. We derived all the equations of motion of this theory and calculated its corresponding transport coefficients. Furthermore, we demonstrated that such transport coefficients are consistent with the linear causality and stability conditions derived in Ref.\ \cite{bd3order}.

Last, we analyzed the derived third-order fluid-dynamical equations within the highly symmetric framework of Bjorken flow. We observed that the currents $\pi^{\mu\nu}$ and $\Theta^{\mu\nu\alpha\beta}$ are the only ones that provide nonvanishing contributions in this flow profile, since we considered massless particles in this comparison. We observed that third-order fluid dynamics derived from the method of moments provides results that are slightly different than a derivation from the Chapman-Enskog method \cite{amaresh}, but are still in good agreement with solutions of the relativistic Boltzmann equation both for LHC and RHIC energies. Nevertheless, the formalism developed throughout this work satisfies causality and stability in the linear regime \cite{bd3order} and, thus, may be solved in more general flow configurations. 

As a future development, we intend to derive a third-order fluid-dynamical theory without resorting to a minimal truncation scheme or to the relaxation time approximation. In this case, instead of directly truncating the moment expansion, one should truncate the moment equation employing a systematic power-counting scheme \cite{dnmr, Denicol:2021}. Such a derivation will provide more accurate expressions for the transport coefficients of our theory.

\section*{Acknoledgments}

C.~V.~P.~B. thanks G. S. Rocha for helpful discussions. C.~V.~P.~B. is funded by Conselho Nacional de Desenvolvimento Científico e Tecnológico (CNPq), process No.~140453/2021-0. G.~S.~D. is funded by CNPq and Fundação de Amparo à Pesquisa do Estado do Rio de Janeiro, process No.~E-26/202.747/2018.

\appendix

\section{List of transport coefficients}
\label{app_coeffs}

In the derivation of the equations of motion of the theory, in Sec.~\ref{sec_30mom}, we have introduced several transport coefficients.

First, in Eq.~\eqref{eom_shear}, we have defined

\begin{align}
\lambda_{\pi\Omega}&=-\left(\frac{\partial\gamma_{-1}^\Omega}{\partial\alpha_0}+h_0^{-1}\frac{\partial\gamma_{-1}^\Omega}{\partial\beta_0}\right),\\
\tau_{\pi\Omega}&=\beta_0\frac{\partial\gamma_{-1}^\Omega}{\partial\beta_0},
\end{align}
where $h_0 = (\varepsilon_0 + P_0)/n_0$. Then, in Eq.~\eqref{eom_rank3} we introduced
\begin{align}
\tau_\Omega&= t_R,\\
\delta_{\Omega\Omega}&=-\frac{1}{3}\tau_\Omega\left(m^2\gamma^\Omega_{-2}+5\right),\\
\ell_{\Omega n}&=-\frac{6}{35}\tau_\Omega\left(m^4\gamma^n_{-2}+5m^2-6\gamma^n_{2}\right),\\
\ell_{\Omega\Omega}&=-\frac{1}{3}\tau_\Omega\left(2m^2\gamma^\Omega_{-2}+7\right),\\
\eta_\Omega&=-\tau_\Omega\left(m^2 \gamma^\pi_{-1}-\gamma^\pi_{1}\right),\\
\lambda_{\Omega\pi}&=-\frac{3}{7}\tau_\Omega\left[m^2\left(\frac{\partial \gamma_{-1}^\pi}{\partial\alpha_0}+h_0^{-1}\frac{\partial \gamma_{-1}^\pi}{\partial\beta_0}\right)-\left(\frac{\partial \gamma_{1}^\pi}{\partial\alpha_0}+h_0^{-1}\frac{\partial \gamma_{1}^\pi}{\partial\beta_0}\right)\right],\\
\lambda_{\Omega\Theta}&=-\tau_\Omega\left(\frac{\partial\gamma_{-1}^\Theta}{\partial\alpha_0}+h_0^{-1}\frac{\partial\gamma_{-1}^\Theta}{\partial\beta_0}\right),\\
\tau_{\Omega \pi} & =  - \frac{3}{7} \tau_\Omega \frac{\beta_0}{\varepsilon_0 + P_0} \left( m^2 \frac{\partial \gamma_{-1}^\pi}{\partial \beta_0} - \frac{\partial \gamma_{1}^\pi}{\partial \beta_0} \right),  \\
\tau_{\Omega\Theta}&=\tau_\Omega\beta_0\frac{\partial\gamma^\Theta_{-1}}{\partial \beta_0},
\end{align}
Finally, in Eq.~\eqref{eom_Theta}, we have
\begin{align}
\tau_\Theta&= t_R,\\
\delta_{\Theta\Theta}&=-\frac{1}{3}\tau_\Theta\left(\gamma^\Theta_{-2}m^2+6\right),\\
\tau_{\Theta\Theta}&=-\frac{4}{11}\tau_\Theta\left(2\gamma^\Theta_{-2}m^2+9\right),\\
\ell_{\Theta\Omega}&=-\frac{4}{9}\tau_\Theta\left(m^2\gamma^\Omega_{-1}-\gamma^\Omega_{1}\right),\\
\ell_{\Theta\pi}&=-\frac{4}{21}\tau_\Theta\left(\gamma^\pi_{-2}m^4+7m^2-8\gamma^\pi_{2}\right),\\
\lambda_{\Theta\Omega}&=-\frac{4}{9}\tau_\Theta\left[m^2\left(\frac{\partial\gamma^\Omega_{-1}}{\partial\alpha_0}+h_0^{-1}\frac{\partial\gamma^\Omega_{-1}}{\partial\beta_0}\right)-\left(\frac{\partial\gamma^\Omega_{1}}{\partial\alpha_0}+h_0^{-1}\frac{\partial\gamma^\Omega_{1}}{\partial\beta_0}\right)\right],\\
\tau_{\Theta\Omega}&=-4\tau_\Theta\left[\gamma^\Omega_{1}-\frac{\beta_0}{9}\left(m^2\frac{\partial\gamma^\Omega_{-1}}{\partial\alpha_0}-\frac{\partial\gamma^\Omega_{1}}{\partial\beta_0}\right)\right].
\end{align}

The remaining transport coefficients that were not explicitly defined here can be found in Ref.~\cite{dnmr}.

In the massless and classical limits, the transport coefficients listed above reduce to
\begin{align}
\lambda_{\pi \Omega} & = - \frac{\beta_0}{28}, \hspace{.1cm}
\tau_{\pi \Omega} = \frac{\beta_0}{7}, \hspace{.1cm}
\delta_{\Omega \Omega} = - \frac{5}{3} \tau_\Omega, \hspace{.1cm} 
\ell_{\Omega n} = \frac{144 \tau_\Omega}{7 \beta^2_0}, \hspace{.1cm}
\tau_{\Omega \Omega} = - \frac{7}{3} \tau_\Omega, \hspace{.1cm}
\eta_\Omega = \frac{6}{\beta_0} \tau_\Omega, \hspace{.1cm}
\lambda_{\Omega \pi} = - \frac{9}{14} \frac{\tau_\Omega}{\beta_0}, \\
\lambda_{\Omega \Theta} & = - \frac{\beta_0 \tau_\Omega}{36}, \hspace{.1cm}
\tau_{\Omega \pi} = - \frac{9 \pi^2}{14} \tau_\Omega n_0, \hspace{.1cm}
\tau_{\Omega \Theta} = \frac{\beta_0 \tau_\Omega}{9}, \hspace{.1cm}
\delta_{\Theta \Theta} = - 2 \tau_\Theta, \hspace{.1cm}
\ell_{\Theta \Theta} = - \frac{36}{11} \tau_\Theta, \hspace{.1cm}
\ell_{\Theta \Omega} = \frac{32 \tau_\Theta}{9 \beta_0}, \\
\ell_{\Theta \pi} & = \frac{64 \tau_\Theta}{\beta^2_0}, \hspace{.1cm}
\lambda_{\Theta \Omega} = - \frac{8}{9} \frac{\tau_\Theta}{\beta_0}, \hspace{.1cm}
\tau_{\Theta \Omega} = - \frac{256 \tau_\Theta}{9 \beta_0},
\end{align}
where we have used that $h_0 = 4/\beta_0$.

\bibliographystyle{apsrev4-1}
\bibliography{refs}

\end{document}